\documentclass[prd,twocolumn,amsmath,showpacs,amssymb,superscriptaddress, floatfix, nofootinbib,10pt]{revtex4}
\usepackage{indentfirst}
\usepackage{tikz}
\usepackage{setspace}
\usepackage{amsfonts,color,xcolor,graphicx,amsfonts,multirow,amsmath}
\usepackage[section]{placeins}
\definecolor{bluee}{rgb}{0,0,1}
\usepackage[dvipdfm, colorlinks=true, colorlinks=true, citecolor=bluee, linkcolor=bluee, urlcolor=bluee]{hyperref}

\usepackage{txfonts}
\DeclareMathAlphabet{\mathcal}{OMS}{cmsy}{m}{n}
\DeclareSymbolFont{largesymbols}{OMX}{cmex}{m}{n}

\definecolor{darkerblue}{rgb}{0.1, 0.1, 0.7}
\definecolor{darkergreen}{rgb}{0.1, 0.5, 0.1}
\definecolor{lightred}{rgb}{1, 0, 0}
\usepackage{ulem}

\allowdisplaybreaks

\usepackage{slashed}
\usepackage{cancel}
\begin{document}
\title{Status of the $D_s^+\to\phi\ell^+\nu_\ell$ decay with a chiral-odd $\phi$-meson light-cone distribution amplitude}
\author{Ya-Xiong Wang$^*$}
\address{Department of Physics, Guizhou Minzu University, Guiyang 550025, P.R.China}
\author{Dan-Dan Hu\footnote{Ya-Xiong Wang and Dan-Dan Hu contributed equally to this work.}}
\address{Department of Physics, Chongqing University, Chongqing 401331, P.R. China}
\author{Wan-Bing Luo}
\address{Department of Physics, Guizhou Minzu University, Guiyang 550025, P.R.China}
\author{Tao Zhong}
\author{Hai-Bing Fu}
\email{fuhb@gzmu.edu.cn}
\address{Department of Physics, Guizhou Minzu University, Guiyang 550025, P.R.China}
\address{Institute of High Energy Physics, Chinese Academy of Sciences, Beijing 100049, P.R.China}

\begin{abstract}
The twist-2 distribution amplitude of the $\phi$-meson has attracted considerable interest due to its unique properties. In this work, we construct the transverse leading-twist light-cone distribution amplitude (LCDA) $\phi_{2;\phi}^\bot(x,\mu_0)$ of the $\phi$-meson using the light-cone harmonic oscillator (LCHO) model, in which a parameter $B_{2;\phi}^\bot$ dominantly control its longitudinal distribution. To explicitly isolate different twist contributions, we employ the right-handed chiral correlator for the QCD light-cone sum rules (LCSR) calculation of $D_s^+\to\phi$ decays, and further, we get the branching fraction,  $\mathcal{B}(D_s^+ \to \phi e^+\nu_e )= (2.271_{-0.243}^{+0.291})\times 10^{-2}$ and
$\mathcal{B}(D_s^+ \to \phi \mu^+\nu_\mu )=(2.250_{-0.240}^{+0.287})\times 10^{-2}$, where errors are squared average of the mentioned error sources. Furthermore, we have extracted the Cabbibo-Kobayashi-Maskawa (CKM) matrix element $|V_{cs}|=0.975_{-0.066}^{+0.067}$ with improved precision through the analysis. Finally, we calculated the polarization parameter and asymmetry parameter for the $D_s^+\to\phi$ decays.
\end{abstract}
\date{\today}

\maketitle

\section{Introduction.}
In current research, studies of charmed mesons, particularly $D_s$ mesons composed of a charm quark and a strange antiquark, have attracted substantially increased attention. The study of $D_s^+\to\phi\ell^+\nu_\ell$ semileptonic decays has attracted widespread attention. This naturally leads to our focused interest in semileptonic decay processes involving the $\phi$-meson.
For a semileptonic decay process, charmed mesons provide useful information about the weak and strong interactions of mesons composed of heavy quarks, which includes the important message on the value of the Cabibbo-Kobayashi-Maskawa (CKM) matrix elements $V_{Qq}$ (with $Q$ representing the heavy quark and $q$ representing the light one).  From a theoretical perspective, semileptonic decays are much simpler than hadronic decays because the final state involves only one meson and one lepton pair. Specifically, the leptonic part can be easily calculated using standard methods, while the hadronic part can be factorized to reduce theoretical uncertainties ~\cite{Cabibbo:1963yz,Kobayashi:1973fv,Wolfenstein:1983yz}. Within the Standard Model (SM), semileptonic decays are ideal for testing lepton flavor universality (LFU) and searching for new physics. The decay $D_s^+\to\phi\ell^+\nu_\ell$ ($\ell=e,\,u$) is highly sensitive to LFU violation, as a deviation between $e$ and $\mu$ modes could signal new physics. Previous studies of $D^{0/+}$, $D_s^+$ mesons~\cite{BESIII:2018nzb, BESIII:2018ccy, Ablikim:2020hsc, BESIII:2019qci, BESIII:2017ikf, Ke:2023qzc} and charm baryons~\cite{BESIII:2021ynj, Belle:2021crz, Belle:2021dgc, BESIII:2015ysy} all agree with the SM within current uncertainties. These decays are described by hadronic form factors, computed using non-perturbative QCD methods.

Before conducting theoretical studies, determining the quark structure of the meson ({\it e.g.} the $\phi$-meson) is essential. The $\phi$-meson, as a vector meson with spin-parity $J^{PC}=1^{--}$, has similar properties to the neutral mesons $\omega$ and $\rho$.
Neutral mesons with hidden-flavor can mix through strong and electromagnetic interactions if they carry identical quantum numbers (e.g. spin, parity, and charge conjugation), since these quantum numbers are strictly conserved in these processes~\cite{Li:2020ylu}.
The tri-meson mixing $\phi-\omega-\rho^0$ system has attracted extensive research efforts, with diverse mixing mechanisms and parameterization approaches being investigated~\cite{Benayoun:2001qz,Benayoun:2007cu,Benayoun:2008tm,Qian:2009dc}.
Within the unified mixing framework of ideal basis, the meson mixing can be described by a tri-mixing matrix, as a detailed in Ref.~\cite{Benayoun:2001qz}.
In the two-meson mixing system, the $\rho-\omega$ mixing can induced either by strong isospin-violating or by electromagnetic effects.
The former is proportional to the mass difference between the $u$, $d$ quarks, {\it i.e.}, $\Delta_{u,d}=m_u - m_d$ and the latter is accompanied by the fine structure constant $\alpha$~\cite{Fritzsch:2001aj}.
In studies of $\rho-\omega$ mixing, various loop mechanisms are employed in different models to achieve the mixing effect, such as
extended Nambu-Jona-Lasinio (NJL) model~\cite{Shakin:1995md},
the global color model~\cite{Mitchell:1996dn},
the hidden local symmetry model~\cite{OConnell:1997ggd},
and the chiral constituent quark model~\cite{Gao:1998gr,Wang:2000gq}.
Meanwhile, several approaches are employed to describe $\rho-\omega$ mixing, including schemes such as ideal mixing (Flavor Eigenstate Symmetry Breaking Models)~\cite{Fritzsch:2000pg,Fritzsch:2001aj}.
Under this mixing scheme, we obtain:
\begin{align}
|\rho_0\rangle &= |\bar{u}u\rangle - |\bar{d}d\rangle
\nonumber\\
&= \cos \alpha |\frac{1}{\sqrt{2}}(\bar{u}u-\bar{d}d)\rangle
-\sin \alpha |\frac{1}{\sqrt{2}}(\bar{u}u+\bar{d}d)\rangle
\\
|\omega\rangle &= |\bar{u}u\rangle + |\bar{d}d\rangle
\nonumber\\
&= \sin \alpha |\frac{1}{\sqrt{2}}(\bar{u}u-\bar{d}d)\rangle
+\cos \alpha |\frac{1}{\sqrt{2}}(\bar{u}u+\bar{d}d)\rangle
\end{align}
The mixing angle $\alpha$ describing the strength of the triplet-singlet mixing is about $-4.1^\circ$, {\it i.e.} a sizeable violation of isospin symmetry is obtained.
Correspondingly, we obtain the mixing parameters: $\cos \alpha=0.997$, $\sin \alpha=-0.071$.
Therefore, neither is the $\rho_0$-meson an isospin triplet, nor is the $\omega$-meson an isospin singlet.
A more detailed discussion can be found in Refs.~\cite{OConnell:1995nse,OConnell:1994czf,
Benayoun:2001qz,OConnell:1996amv,Li:2019xwh}.
In addition, $\omega-\phi$ mixing is well-established for the $\phi$-meson~\cite{Benayoun:1999fv,Kucukarslan:2006wk,Benayoun:2007cu,
Gronau:2008kk,Gronau:2009mp}. Meson mixing is an intriguing phenomenon that can explain certain decay processes of heavy mesons. In this study, we aim to clarify the impact of $\omega-\phi$ mixing on the flavor wave function (WF) components of the $\phi$-meson. Within the framework of the singlet-octet mixing scheme, the $\omega$ and $\phi$-mesons are mixtures of the SU(3) singlet $\omega_0$ state and octet $\omega_8$ state~\cite{Yang:2025qgh}:
\begin{align}
\left( \begin{array}{c}
\phi\\
\omega\\
\end{array}\right)=
\left(\begin{matrix}
\mathrm{cos}\theta_V& -\mathrm{sin}\theta_V\\
\mathrm{sin}\theta_V& \mathrm{cos}\theta_V\\
\end{matrix}\right)
\left(\begin{array}{c}
\omega_8\\
\omega_0\\
\end{array}\right),
\end{align}
where $\omega_8=(u\bar{u}+d\bar{d}-2s\bar{s})/\sqrt{6}$ and $\omega_0=(u\bar{u}+d\bar{d}+s\bar{s})/\sqrt{3}$. Alternatively, in the quark flavor basis mixing scheme, the physical states of $\omega$ and $\phi$-mesons can be further expressed as:
\begin{align}
\label{eq:phi}
&|\phi\rangle = \mathrm{sin}\varphi_V|\omega_q\rangle-\mathrm{cos}\varphi_V|\omega_s\rangle,
\\
&|\omega\rangle = \mathrm{cos}\varphi_V|\omega_q\rangle+\mathrm{sin}\varphi_V|\omega_s\rangle,
\end{align}
where $|\omega_q\rangle=\frac{1}{\sqrt{2}}(u\bar{u}+d\bar{d})$ and $|\omega_s\rangle=s\bar{s}$ are the quark flavor bases.
The mixing angle $\varphi_V$ crucially determines the flavor composition.
The value of the mixing angle $\varphi_V=3.4\pm0.3^\circ$ is derived within the framework of chiral perturbation theory~\cite{Kucukarslan:2006wk}.
The KLOE Collaboration determined the $\omega-\phi$ mixing angle $\varphi_V=3.32\pm0.09^\circ$ by fitting their experimental measurements~\cite{Ambrosino:2009sc}.
In studying $\chi_{cJ} \to \phi \phi $ decays, the mixing angle $\varphi_V=3.7^\circ$ can be extracted by applying the Gell-Mann-Okubo mass formula~\cite{Huang:2021kfm}.
The mixing angle $\varphi_V=3.1\pm0.3^\circ$ can be predicted within the framework of the Nambu-Jona-Lasinio model~\cite{Volkov:2020jor}.
A detailed discussion on this can be found in Refs.~\cite{Bramon:1994cb,Bramon:1994pq,
Benayoun:2000ti}.
In current research, the value of $\cos\varphi_V$ remains close to 1.
As can be seen in Eq.~\eqref{eq:phi}, the second term dominates, indicating that the $\phi$-meson is predominantly composed of an $s\bar{s}$ quark pair. Therefore, in subsequent studies, we treat the neutral vector $\phi$-meson as a pure $s\bar{s}$ quarkonium state.

Recent experimental studies of $D_s^+$ decays to $s\bar{s}$-containing final states have achieved significant progress, with notable advances in precision measurements and new channel observations. The latest analyses focus on improved determinations of form factors, polarization parameters, and tests of lepton flavor universality through these decay modes. These developments provide stringent constraints on QCD dynamics and potential beyond SM effects.
In 2023, the BESIII collaboration reported a precision measurement of $\mathcal{B}(D_s^+\to \phi\mu^+\nu_\mu)=(2.25\pm 0.09\pm 0.07)\times10^{-2}$ ~\cite{BESIII:2023opt}.
In 2017, the BESIII collaboration reported the first measurements of branching fractions for $D_s^+ \to \phi e^+\nu_e$ and $D_s^+ \to \phi \mu^+\nu_\mu$ decays, obtaining $\mathcal{B}(D_s^+\to\phi e^+\nu_e)=(2.26\pm 0.45\pm 0.09)\times 10^{-2}$, $\mathcal{B}(D_s^+\to\phi \mu^+\nu_\mu)=(1.94\pm0.53\pm0.09)\times 10^{-2}$, which demonstrated consistency with lepton flavor universality within uncertainties~\cite{BESIII:2017ikf}.
Additionally, the PDG has incorporated data from the BarBar collaboration~\cite{BaBar:2008gpr} and the CLEO collaboration~\cite{Hietala:2015jqa}. This conclusively demonstrates the essential importance of studying semileptonic decays of charmed mesons.

On the theoretical front, numerous non-perturbative methods are currently being employed to investigate the decay processes $D_s^+\to\phi\ell^+\nu_\ell$ with $\ell=(e,\mu)$. Several groups have also presented results, {\it i.e.} the heavy quark effective field theory (HQEFT)~\cite{Wu:2006rd}, the heavy meson chiral Lagrangians (HM$\chi$T)~\cite{Fajfer:2005ug}, the covariant quark model (CQM)~\cite{Melikhov:2000yu,Soni:2019huk}, the traditional three-point sum rules (3PSR)~\cite{Du:2003ja,Bediaga:2003hr}, the covariant light-front quark model (CLFQM)~\cite{Verma:2011yw,Cheng:2017pcq}, the light-front quark model (LFQM)~\cite{Chang:2019mmh}, the light-cone sum rules (LCSR)~\cite{Aliev:2004vf}, the covariant confining quark model (CCQM)~\cite{Ivanov:2019nqd}, the lattice QCD (LQCD)~\cite{Donald:2013pea,Donald:2011ff,Koponen:2011ev}, the relativistic quark model (RQM)~\cite{Faustov:2019mqr,Faustov:2013ima,Faustov:2012mt,Ebert:2009ua} and the chiral unitary approach ($\chi$UA)~\cite{Hussain:1995jq,Sekihara:2015iha}. From 2000 to 2019, theoretical predictions for the branching fractions $\mathcal{B}(D_s^+\to\phi\ell^+\nu_\ell)$ consistently exhibited central values that deviated from experimental measurements by varying degrees. Specifically, the theoretical calculations for the branching fraction of $D_s^+\to\phi e^+\nu_e$ yield central values ranging between $1.8\times 10^{-2}$ and $3.1 \times 10^{-2}$, while the central values for the $D_s^+\to\phi \mu^+\nu_\mu$ branching fraction calculations fall between $2.4 \times10^{-2}$ and $2.9 \times10^{-2}$. The HQEFT provides a unified framework for treating quark and antiquark fields symmetrically in heavy-light mesons like $D_{(s)}$. In studies of $D_s\to\phi$ decays, it predicts branching fractions of $\mathcal{B}(D_s^+\to\phi e^+\nu_e)=2.53_{-0.40}^{+0.37}\times10^{-2}$ and $\mathcal{B}(D_s^+\to\phi\mu^+\nu_\mu)=2.40_{-0.40}^{+0.35}\times10^{-2}$~\cite{Wu:2006rd}. Within the CLFQM, the branching fractions of $D_s^+\to\phi$ decays have been investigated through improved treatments of both the strange axial vector meson masses and mixing angles, yielding results of $\mathcal{B}(D_s^+\to\phi e^+\nu_e)= (3.1\pm0.3)\times 10^{-2}$ and $\mathcal{B}(D_s^+\to\phi \mu^+\nu_\mu)= (2.9\pm0.3)\times 10^{-2}$~\cite{Verma:2011yw}. From the perspective of transition form factors (TFFs), the complete set of axial and vector form factors for the $D_s\to \phi\ell^+\nu_\ell$ decay from full LQCD is determined, in Ref.~\cite{Donald:2013pea}. The valence quarks are implemented using the Highly Improved Staggered Quark action and normalise the appropriate axial and vector currents fully nonperturbatively. Simultaneously, the theoretical result of TFFs was obtained, {\it i.e.} $A_1(0)=0.615(24)$, $A_2(0)=0.457(78)$, $A_0(0)=0.706(37)$ and $V(0)=1.059(124)$. Through the joint analysis of experimental measurements and theoretical predictions, it becomes evident that the $D_s^+\to \phi \ell^+\nu_\ell$ decay process still requires further exploration to achieve satisfactory consistency between theory and experiment.

In order to achieve a precise standard model (SM) determination on the decay $D_s^+\to \phi \ell^+\nu_\ell$, we need to determine the nonperturbative hadronic matrix elements, or equivalently, the $D_s^+\to \phi$ TFFs  $A_1(q^2)$, $A_2(q^2)$, $V(q^2)$ and $A_0(q^2)$, where $V(q^2)$ is the TFF defined via a vector current, $A_{1,2,0}(q^2)$ are the TFFs defined via an axial-vector current~\cite{Fu:2014uea}.
In this paper, one have selected the QCD LCSR approach subsequent calculations. This method is applicable in the low and intermediate $q^2$-regions, and can be further extrapolated to entire allowed $q^2$ range. The LCSR are based on the operator product expansion (OPE) near the light cone, where the non-perturbative dynamics are parameterized by leading-twist light-cone distribution amplitude (LCDA) of increasing twists. Therefore, more precise LCSR predictions with reduced theoretical uncertainties will significantly enhance the study of these TFFs and the $D_s^+ \to \phi \ell^+\nu_\ell$ decays. Vector meson LCDAs posses intricate structures that can be systematically organized through the parameter $\delta\simeq m_\phi/m_c\sim 0.80$. Following the methodology outlined in Ref.~\cite{Ball:2004rg}, one present the complete set of $\phi$-meson LCDAs across different twist-structures in Table~\ref{Tab:Delta}, where the subscripts 2,3,4 stand for the twist-2, the twist-3 and the twist-4 LCDAs, respectively.

\begin{table}[t]
\footnotesize
\begin{center}
\caption{The $\phi$-meson LCDAs with different twist-structures up to $\delta^3$-order, where $\delta\simeq m_\phi/m_c$.}
\label{Tab:Delta}
\begin{tabular}{c c c c}
\hline
~~~~~~~~~~~~~~~&twist-2
&twist-3
&twist-4
\\
\hline
$\delta^0$
&$\phi_{2;\phi}^\bot$
&/
&/
\\
$\delta^1$
&$\phi_{2;\phi}^\parallel$
&$\phi_{3;\phi}^\bot$,$\psi_{3;\phi}^\bot$,$\Phi_{3;\phi}^\parallel$,$\tilde{\Phi}_{3;\phi}^\parallel$                                              &/
\\
$\delta^2$
&/
&$\phi_{3;\phi}^\parallel$,$\psi_{3;\phi}^\parallel$,$\Phi_{3;\phi}^\bot$                                             &$\phi_{4;\phi}^\bot$,$\psi_{4;\phi}^\bot$,$\Psi_{4;\phi}^\bot$,$\tilde{\Psi}_{4;\phi}^\bot$
\\
$\delta^3$
&/
&/                                             &$\phi_{4;\phi}^\parallel$,$\psi_{4;\phi}^\parallel$
\\
\hline
\end{tabular}
\end{center}
\end{table}

Although these higher-twist LCDAs are suppressed at $\delta^1$-order or beyond, they may still contribute significantly to LCSR calculations. Therefore, actively constraining these uncertainty sources is essential for achieving more precise LCSR predictions. In this paper, we have for the first time constructed the leading-twist transverse LCDA $\phi_{2;\phi}^{\bot}(x,\mu)$ of the $\phi$-meson using the light cone harmonic oscillator (LCHO) model. The LCHO model is constructed based on the Brodsky-Huang-Lepage (BHL) prescription and Melosh-Wigner transformation, where the Melosh-Wigner transformation connects light-cone spin states with conventional instant-form spin wave functions. This transformation serves as one of the most essential components in light-cone formalization~\cite{Melosh:1974cu}. In this context, we can utilize the Melosh-Wigner rotation~\cite{Yu:2007hp} to construct the light-cone wave functions for quark-antiquark Fock states within the light-cone quark model. The complete light-front wave function is achieved through the evaluation of both spin and momentum-space wave functions.
The two-particle Fock state expansion of the $\phi$-meson includes both longitudinal and transverse types of spin configurations. Each configuration corresponds to distinct quark-antiquark helicity combinations.
The construction utilizes a model parameter that dominantly governs its transverse behavior $\phi_{2;\phi}^{\bot}(x,\mu_0)$ within well-defined ranges.
Furthermore, to isolate the influence of high-twist LCDAs, we employ the right-handed chiral correlator~\cite{Wu:2025kdc} in the LCSR calculations. By employing these LCSRs, we have conducted comprehensive comparisons with various experimental and theoretical groups to determine the decay branching fractions $\mathcal{B}(D_s^+\to\phi\ell^+\nu_\ell)$ and CKM matrix element $|V_{cs}|$.

The rest of this paper is organized as follows: In Sec.~\ref{sec:framework}, we describe the computational techniques used to derive the $D_s\to\phi$ and provide the relevant expressions for the decay widths. In Sec.~\ref{sec:analysis}, we explain the construction of the LCHO model and the determination of its input parameters, along with a detailed numerical analysis and discussion. Section~\ref{sec:summary} is reserved for a summary.

\section{Theoretical Framework}\label{sec:framework}
In SM framework, the weak interaction governing the $D_s^+\to\phi$ semileptonic decay is dominated by the vector-axial vector current. According to the definition of the current, the TFFs of $D_s\to\phi$ can be divided into the vector current contribution $V(q^2)$ and the axial-vector current contributions $A_{1,2,0}(q^2)$, which together describe the dynamics of the hadronic matrix element. These TFFs are coming from the transition matrix element within the current $\bar{s}(x)\gamma_\mu\Gamma_5^{\rm L}c(x)$ with $\Gamma_5^{\rm L}=(1-\gamma_5)$, for which we adopt the following matrix element to derive the $D_s^+\to\phi$ TFFs:
\begin{align}
&\langle \phi(p,\lambda)|\overline{s}\gamma_\mu \Gamma_5^{\rm L} c | D_s^+(p+q) \rangle = -e_\mu ^{*(\lambda)}(m_{D_s^+} + m_{\phi}) A_1(q^2)
\nonumber\\
&\qquad + i(e^{*(\lambda)} \cdot q) \frac{(2p + q)_\mu }{m_{D_s^+}+m_{\phi}} A_2(q^2)+iq_\mu (e^{*(\lambda)} \cdot q)\frac{2m_{\phi}}{q^2}  \nonumber\\
&\qquad \times[ A_3(q^2)- A_0(q^2) ]+ \epsilon_{\mu\nu\alpha\beta} e^{*(\lambda)\nu} q^{\alpha} p^{\beta} \frac{2V(q^2)}{m_{D_s^+} + m_{\phi}},
\end{align}
where $e^{(\lambda)}$ stands for the $\phi$-meson polarization vector with $\lambda$ being its transverse $(\bot)$ or longitudinal $(\parallel)$ component, respectively. $p$ is the $\phi$-meson momentum and $q= (p_{D_s^+}-p)$ is the momentum transfer between the $D_s^+$-meson and the $\phi$-meson. There are some relations among those TFFs, that not all of them are independent as follows,
\begin{align}
A_3(q^2)&=\frac{m_{D_s^+}+m_{\phi}}{2m_{\phi}}A_1(q^2)-\frac{m_{D_s^+}-m_{\phi}}{2m_{\phi}}A_2(q^2),
\end{align}
And at the large recoil point $q^2=0$, we have
\begin{align}
A_0(0)&=A_3(0).
\end{align}
To obtain the LCSRs for those TFFs, we currently process the following correlator:
\begin{align}
\hspace{-0.3cm}\mathrm{\Pi}_\mu(p,q) =i \! \int \! d^4x e^{iq\cdot x} \langle \phi(p,\lambda)|T\{ \bar{s}(x)\gamma_\mu\Gamma_5^{\rm L}c(x), j^\dagger_{D_s^+}(0) \}|0  \rangle.
\end{align}
For the current $j^\dagger_{D_s^+}(x)$ selection, the adoption of chiral current serves~\cite{Huang:2001xb} this purpose by suppressing the in a complex series involving all possible $\phi$-meson twist-structures. Therefore, the chiral current can be selected as $j^\dagger_{D_s^+}(x)= i m_c\bar{c}(x)\Gamma_5^{\rm R}q(x)$ with $\Gamma_5^{\rm R} = (1+\gamma_5)$ to proceed with subsequent calculations. This advantage manifests in the ability to highlight different twist contributions of $\phi$-meson LCDAs to the TFFs through judicious selection of chiral current.   It is particularly noteworthy that the hadronic representation of this correlator encompasses not only the standard $J^P=0^-$ resonances but also extra $J^P=0^+$ resonant states. This represents the price when introducing chiral currents into LCSR calculations.

The correlators are analytic $q^2$-functions defined at both the time-like and the space-like $q^2$-region. On the one hand, in the time-like $q^2$-region, the long-distance quark-gluon interactions become important and, eventually, the quarks form hadrons. In this region, one can insert a complete series of intermediate hadronic states in the correlator and obtain its hadronic representation by isolating the pole term of the lowest pseudoscalar $D_s^+$-meson:
\begin{align}
\Pi_\mu ^{\rm H}(p,q)&=\frac{\langle \phi|\bar{s}\gamma_\mu \Gamma_5^L c |D_s^+ \rangle \langle  D_s^+| \bar{c}im_c\gamma _5q_1 |0 \rangle }{m_{D_s^+} ^2 -(p+q)^2}
\nonumber\\
& +\sum_{\rm H}{\frac{\langle  \phi| \bar{s}\gamma _\mu \Gamma_5^L c |{D_s^+}^{\rm H} \rangle \langle  {D_s^+}^{\rm H}| \bar{c}im_c \Gamma_5^R q_1 |0 \rangle }{m_{{D_s^+}^{\rm H}}^2 -(p+q)^2}}
\end{align}
which can separated by the four Lorentz structure with the following form
\begin{align}
\Pi_\mu ^{\rm H}(p,q)&=e_\mu ^{*(\lambda)}\Pi_1^{\rm H} + (e_\mu ^{*(\lambda)}\cdot q) (2p+q)_\mu \Pi_2^{\rm H}+(e_\mu ^{*(\lambda)}\cdot q)q_\mu \Pi_3^{\rm H}
\nonumber\\
&  + i \epsilon_\mu ^{\nu\alpha\beta} e_{\nu}^{*(\lambda)}q_{\alpha}p_{\beta} \Pi_{4}^{\rm H}
\end{align}
where $\langle  D_s^+ |\bar{c}im_c\gamma _5q_1 |0 \rangle = m_{D_s^+}^2 f_{D_s^+}$ with $f_{D_s^+}$ standing for the ${D_s^+}$-meson decay constant. The invariant amplitudes $\Pi_{1,2,3,4}^{\rm H}$ are respectively
\begin{align}
\Pi_{1}^{\rm H}&[q^2 ,(p+q)^2 ]=
\frac{m_{D_s^+}^2 f_{D_s^+}(m_{D_s^+}+m_{\phi})}{m_{D_s^+}^2 -(p+q)^2 }A_{1}(q^2 )
\nonumber\\
&\hspace{1.2cm}+\int_{s_{0}}^{\infty}\frac{\rho_{1}^{\rm H}}{s-(p+q)^2 }ds+\text{subtractions}
\\
\Pi_{2}^{\rm H}&[q^2 ,(p+q)^2 ]  =\frac{m_{D_s^+}^2 f_{D_s^+}A_{2}(q^2 )}{(m_{D_s^+}+m_{\phi})[m_{D_s^+}^2 -(p+q)^2 ]}
\nonumber\\
&\hspace{1.2cm}+\int_{s_{0}}^{\infty}\frac{\rho_{2}^{\rm H}}{s-(p+q)^2 }ds
+\text{subtractions}
\\
\Pi_{3}^{\rm H}&[q^2 ,(p+q)^2 ]
=\frac{2m_{D_s^+}^2 f_{D_s^+}m_{\phi}[A_{3}(q^2 )-A_{0}(q^2 )]}{q^2 [m_{D_s^+}^2 -(p+q)^2 ]}
\nonumber\\
&\hspace{1.2cm}+\int_{s_{0}}^{\infty}\frac{\rho_{3}^{\rm H}}{s-(p+q)^2 }ds+\text{subtractions}
\\
\Pi_{4}^{\rm H}&[q^2 ,(p+q)^2 ]
=\frac{2m_{D_s^+}^2 f_{D_s^+}V(q^2 )}{(m_{D_s^+}+m_{\phi})[m_{D_s^+}^2 -(p+q)^2 ]}
\nonumber\\
&\hspace{1.2cm}+\int_{s_{0}}^{\infty}\frac{\rho_{4}^{\rm H}}{s-(p+q)^2 }ds+\text{subtractions}
\end{align}
Furthermore, one can employ dispersion relations to effectively substitute the contributions from both higher resonances and the continuum states. The spectral densities $\rho_{1,2,3,4}^{\rm H}$ can be approximated by applying the conventional quark-hadron duality ansatz
\begin{align}
\rho_{1,2,3,4}^{\rm H}=\rho_{1,2,3,4}^{\mathrm{QCD}}\theta(s-s_0).
\end{align}
On the other hand, in the space-like $q^2$-region, the correlator can be calculated using QCD OPE. Here, the condition $(p+q)^2-m_c^2 \ll 0$ with momentum transfer $q^2\sim \mathcal{O} \ll m_c^2$ corresponds to sufficiently small light-cone distances $x^2\rightsquigarrow 0$, thereby ensuring the validity of the OPE. The full $c$-quark propagator states
\begin{align}
&\langle 0|\mathrm{T}\{c(x)\bar{c}(0)\}|0\rangle \! = \! i \! \int\! \frac{d^4k}{(2\pi)^4} e^{-ik\cdot x} \bigg\{ \! \frac{\cancel{k}+m_c}{m_{c}^{2}-k^2}\! - \! g_s \! \int_0^1 \! dv
\nonumber \\
&\qquad\times G^{\mu \nu}(vx)\left[ \frac{1}{2}\frac{\cancel{k}+m_c}{(m_{c}^{2}-k^2)^2}\sigma _{\mu \nu}+\frac{v}{m_{c}^{2}-k^2}x_{\mu}\gamma _{\nu} \right] \bigg\} ,
\end{align}
where $G_{\mu\nu}$ is the gluonic field strength and $g_s$ denotes the strong coupling constant. Utilizing this $c$-quark propagator and performing the OPE on the correlator, we obtain the QCD expansion of $\Pi_\mu^\mathrm{QCD}$ incorporating both two-particle and three-particle Fock state contributions.

\begin{widetext}
\begin{align}
\Pi_\mu ^{\mathrm{QCD}}
&=m_c \int\frac{d^{4}xd^{4}k}{(2\pi)^{4}}\,e^{i(q-k)\cdot x}\left\{\frac{1}{m_c ^2 -k^2 }\, \right\{2k^{\mu}\,\langle \phi(p,\lambda)|\bar{s}(x)q_{1}(0)|0\rangle\,-\,2ik^{\nu}\langle \phi(p,\lambda)|\bar{s}(x)\sigma_{\mu\nu}q_{1}(0)|0\rangle
\nonumber\\
&\hspace{0.35cm}-\epsilon_{\mu\nu\alpha\beta}k^{\nu}\langle \phi(p,\lambda)|\bar{s}(x)\sigma_{\alpha\beta}q_{1}(0)|0\rangle\!\bigg\}-\int\! dv\bigg\{\frac{k^{\nu}}{(m_c ^2 -k^2 )^2 }\!\bigg[\!-i\langle \phi(p,\lambda)|\bar{s}(x)g_{s}G_{\mu\nu}(vx)q_{1}(0)|0\rangle\bigg]
\nonumber \\
&\hspace{0.35cm}-2\langle \phi(p,\lambda)\,|\,\bar{s}(x)\sigma_{\mu\alpha}g_sG^{\alpha\nu}(vx)q_1(0)\,|\,0\rangle
\!+\!2\,i\,\langle \phi(p,\lambda)\,|\,\bar{s}(x)ig_s\widetilde{G}_{\mu\nu}(vx)\gamma_5q_1(0)\,|\,0\rangle\bigg]
\!+\!\frac{2vx_{\alpha}}{m_c ^2 -k^2 }
\nonumber\\
&\hspace{0.35cm}\times\left[-\langle \phi(p,\lambda)|\bar{s}(x)g_{s}G_{\mu\alpha}(vx)q_{1}(0)|0\rangle-i\langle \phi(p,\lambda)|\bar{s}(x)\sigma_{\mu\beta}g_{s}G^{\alpha\beta}(vx)q_{1}(0)|0\rangle\right]\bigg\}
\bigg\}
\end{align}
where $\widetilde{G}_{\mu\nu}(vx)=\epsilon_{\mu\nu\alpha\beta}G^{\alpha\beta}(vx)/2$. In this work, we employ right-handed chiral correlators in the LCSR calculations to highlight the contributions from the transverse DAs of the $\phi$-meson, consequently adopting chiral-odd basis DAs. Distinct from chiral-even DAs, chiral-odd DAs serve as an effective probe for symmetry breaking. Up to twist-4 accuracy, the non-zero meson-to-vacuum matrix elements with various $\gamma$-structures, {\it i.e.} $\Gamma=1$, $i\gamma_5$ and $\sigma_{\mu\nu}$, adopt the following form~\cite{Ball:2007zt}:
\begin{align}
\langle \phi(p,\lambda)|\bar{s}(x)\sigma_{\mu\nu}q_{1}(0)|0\rangle
&=-if_{\phi}^{\perp}\!\int_{0}^{1}\!due^{iu(p\cdot x)}\bigg\{(e_\mu ^{*(\lambda)}p_{\nu}\!-\!e_{\nu}^{*(\lambda)}p_\mu )\bigg[\phi_{2;\phi}^{\perp}(u)
\!+\!\frac{m_{\phi}^2 x^2 }{16}\phi_{4;\phi}^{\perp}(u)\bigg]
\nonumber\\
&\hspace{0.35cm}+(p_\mu x_{\nu}-p_{\nu}x_\mu )\,\frac{e^{*(\lambda)}\cdot x}{(p\cdot  x)^2 }m_{\phi}^2 \left[\phi_{3;\phi}^{\parallel}(u)-\frac{1}{2}\phi_{2;\phi}^{\perp}(u)
-\frac{1}{2}\psi_{4;\phi}^{\perp}
(u)\right]
\nonumber\\
&\hspace{0.35cm}+\frac{1}{2}\left[e_\mu ^{*(\lambda)}x_{\nu}
-e_{\nu}^{*(\lambda)}x_\mu \right]\frac{m_{\phi}^2 }
{p\cdot x}\left[\psi_{4;\phi}^{\perp}(u)-\phi_{2;\phi}^{\perp}(u)\right]\bigg\}
\label{eq:twist2twoparticle}
\\
\langle \phi(p,\lambda)|\bar{s}(x)q_1(0)|0\rangle
&=-\frac{i}{2}f_{\phi}^{\perp}\left[e^{*(\lambda)}\cdot x\right]m_{\phi}^2 \int_{0}^{1}due^{iu(p\cdot x)}\psi_{3;\phi}^{\parallel}(u)
\label{eq:twist3twoparticle}
\\
\langle \phi(p,\lambda) |\bar{s} (x) \sigma_{\alpha\beta} g_{s} G^{\mu\nu}(vx) q_{1}(0) | 0 \rangle
&= m_{\phi}^2 f_{\phi}^{\perp}\frac{e^{*(\lambda)}\cdot x}{2(p\cdot x)}\left[p_\mu \left(p_{\alpha}g_{\beta\nu}^{\perp}-p_{\beta}g_{\alpha\nu}^{\perp}\right)
-p_{\nu}\left(p_{\alpha}g_{\beta\mu}^{\perp}-p_{\beta}g_{\alpha\mu}^{\perp}\right)\right] \Phi_{3;\phi}^{\perp}(v,p\cdot x)
\label{eq:threeparticle1}
\\
\langle \phi(p,\lambda)|\bar{s}(x)g_sG^{\mu\nu}(vx)q_1(0)|0\rangle
&=-im_{\phi}^2f_{\phi}^{\perp}\left[e_{\perp\mu}^{*(\lambda)}
p_{\nu}-e_{\perp\nu}^{*(\lambda)}p_\mu \right]\Psi_{4;\phi}^{\perp}(v,p\cdot x)
\label{eq:threeparticle2}
\\
\langle \phi(p,\lambda)|\bar{s}(x)ig_s\tilde{G}_{\mu\nu}(vx)\gamma_5q_1(0)|0\rangle
&=im_{\phi}^2f_{\phi}^\perp\left[e_{\perp\mu}^{*(\lambda)}p_{\nu}
-e_{\perp\nu}^{*(\lambda)}p_\mu \right]\tilde{\Psi}_{4;\phi}^\perp(v,p\cdot x)
\label{eq:threeparticle3}
\end{align}
\end{widetext}
where $f_\phi^\perp$ represents the $\phi$-meson decay constant,
\begin{align}
\langle \phi(p,\lambda)|\bar{s}(0)\sigma_{\mu\nu}q_1(0)|0\rangle
=if_{\phi}^\perp\left[e_\nu^{(\lambda)}p_\mu-e_\mu^{(\lambda)}p_\nu\right]
\nonumber
\end{align}
where we have adopted the following standard conventions:
\begin{align}
&g_{\mu\nu}^\perp =g_{\mu\nu}-\frac{p_\mu x_\nu+p_\nu x_\mu}{p\cdot x}
\nonumber\\
&e_\mu ^{\lambda}=\frac{e^{\lambda}\cdot x}{p\cdot x}\left[p_\mu -\frac{m_{\phi}^2 }{2(p\cdot x)}x_\mu \right]+e_{\perp\mu}^{\lambda}
\nonumber\\
& \phi(v,p\cdot x)=\int\mathcal{D}\alpha e^{ipx(\alpha_{1}+v\alpha_{3})}K(\underline{\alpha})
\nonumber
\end{align}
Where $\mathcal{D}\alpha = d\alpha_1 d\alpha_2 d\alpha_3 \delta(1-\alpha_1-\alpha_2-\alpha_3)$ and $K(\underline{\alpha})$ stands for the twist-3 or twist-4 DA $\Phi^\perp_{3;\phi}(\underline{\alpha})$, $\Psi^\perp_{4;\phi}(\underline{\alpha})$ or $\tilde{\Psi}^\perp_{4;\phi}(\underline{\alpha})$, in which $\underline{\alpha}={\alpha_1,\alpha_2,\alpha_3}$ corresponds to the momentum fractions carried by the antiquark, quark and gluon, respectively.

Subsequently, by equating the correlators in different $q^2$-regions and applying the conventional Borel transformation, one obtain the required LCSRs for the $D_s^+ \to \phi$ TFFs, {\it i.e.}
\begin{widetext}
\begin{align}
A_1(q^2)&=\frac{m_{b}m_{\phi}^2 f_{\phi}^{\perp}}{f_{D_s^+}m_{D_s^+}^2 (m_{D_s^+}+m_{\phi})}
\bigg\{\int_{0}^{1}\frac{du}{u}
\,e^{\frac{m_{D_s^+}^2 -s(u)}{M^2 }}\,\bigg\{\,\frac{\mathcal{C}}{um_{\phi}^2 }\,\Theta(c(u,s_{0}))\,\,
\phi_{2;\phi}^{\perp}(u,\mu)+\Theta(c(u,s_{0}))\,\,\psi_{3;\phi}^{\parallel}(u)-\frac{1}{4}
\nonumber\\
&\hspace{0.42cm}\times\left[\frac{m_c ^2 \mathcal{C}}{u^{3}M^{4}}\widetilde{\widetilde{\Theta}}(c(u,s_{0}))
\!+\!\frac{\mathcal{C}-2m_c ^2 }{u^2 M^2 }\widetilde{\Theta}(c(u,s_{0}))
\!-\!\frac{1}{u}\Theta(c(u,s_{0}))\right]\phi_{4;\phi}^{\perp}(u)
-2\bigg[\frac{\mathcal{C}}{u^2 M^2 }\widetilde{\Theta}(c(u,s_{0}))\!-\!\frac{1}{u}
\nonumber\\
&\hspace{0.42cm}\times\!\Theta(c(u,s_{0}))\bigg]I_{L}(u)
\!-\!\biggl[\frac{2m_c ^2 }{uM^2 }\widetilde{\Theta}(c(u,s_{0}))
\!+\!\Theta(c(u,s_{0}))\biggr]H_{3}(u)\biggr\}
\!+\!\!\int\!\mathcal{D}\alpha_{i}\!\int_{0}^{1}dve^{\frac{m_{D_s^+}^2 -s(X)}
{M^2 }}\bigg[\frac{\mathcal{C}}{2X^{3}M^2 }
\nonumber\\
&\hspace{0.42cm}-\frac{1}{2X^2 }\bigg]\Theta(c(X,s_{0}))\left[(4v-1)\Psi_{4;\phi}^{\perp}(\underline{\alpha})
-\widetilde{\Psi}_{4;\phi}^{\perp}(\underline{\alpha})\right]\bigg\}
\\
A_2(q^2)&=\frac{m_c (m_{D_s^+}+m_{\phi})m_{\phi}^2 f_{\phi}^{\perp}}{f_{D_s^+}m_{D_s^+}^2 }
\,\Bigg\{\,\int_{0}^{1}\frac{du}{u}e^{\frac{m_{D_s^+}^2 -s(u)}
{M^2 }}\,\Bigg\{\,\frac{1}{m_{\phi}^2 }\,\Theta(c(u,s_{0}))\,\phi_{2;\phi}^{\perp}(u,\mu)\,-\,\frac{1}{M^2 }
\,\widetilde{\Theta}(c(u,s_{0}))\,\psi_{3;\phi}^{\parallel}(u)
\nonumber\\
&\hspace{0.42cm}-\!\frac{1}{4}\!\left[\frac{m_c ^2 }{u^2 M^{4}}\widetilde{\widetilde{\Theta}}(c(u,s_{0}))
\!+\!\frac{1}{uM^2 }\widetilde{\Theta}(c(u,s_{0}))\right]\!\phi_{4;\phi}^{\perp}(u)
\!+\!2\!\left[\frac{\mathcal{C}-2m_c ^2 }{u^2 M^{4}}\widetilde{\widetilde{\Theta}}(c(u,s_{0}))\
\!-\!\frac{1}{uM^2 }\widetilde{\Theta}(c(u,s_{0}))\!\right]
\nonumber\\
&\hspace{0.42cm}\times \!I_{L}(u)\!-\!\frac{1}{M^2 }\widetilde{\Theta}(c(u,s_{0}))H_{3}(u)\biggr\}
\!+\!\int\mathcal{D}\alpha_{i}\int_{0}^{1}dve^{\frac{m_{D_s^+}^2 -s(X)}
{M^2 }}\!\frac{1}{2X^2 M^2 }\Theta(c(X,s_{0}))\biggl[(4v-1)\Psi_{4;\phi}^{\perp}
(\underline{\alpha})\biggr]
\nonumber\\
&\hspace{0.42cm}-\widetilde{\Psi}_{4;\phi}^{\perp}(\underline{\alpha})
+4v\Phi_{3;\phi}^{\perp}(\underline{\alpha})\Biggr]\Biggr\}
\\
V(q^2 )&=\frac{m_c (m_{D_s^+}+m_{\phi})f_{\phi}^{\perp}}{f_{D_s^+}m_{D_s^+}^2 }
\int_{0}^{1}\frac{du}{u}e^{\frac{m_{D_s^+}^2 -s(u)}{M^2 }}\biggl\{\,
\Theta(c(u,s_{0}))\,\phi_{2;\phi}^{\perp}(u,\mu)-\frac{m_{\phi}^2 }{4}
\biggl[\frac{m_c ^2 }{u^2 M^{4}}\,\widetilde{\widetilde{\Theta}}(c(u,s_{0}))
+\frac{1}{uM^2 }\,\biggr]\,\biggr]
\nonumber\\
&\hspace{0.42cm}\times\widetilde{\Theta}(c(u,s_{0}))\biggr]\phi_{4;\phi}^{\perp}(u)\biggr\}
\\
A_3(q^2)&-A_0(q^2)=\frac{m_c m_\phi f_{\phi}^{\perp} q^2}{2 f_{D_s^+} m_{D_s^+}^2}
\bigg\{\! \int_{0}^{1}\frac{du}{u}e^{\frac{m_{D_s^+}^2 -s(u)}{M^2 }} \bigg\{\!-\frac{1}{m_\phi^2}
\Theta(c(u,s_{0}))\phi_{2;\phi}^{\perp}(u,\mu)- \frac{2-u}{u M^2} \widetilde{\Theta}(c(u,s_{0}))
\psi_{3;\phi}^{\parallel}(u)
\nonumber\\
&\hspace{0.42cm} + \frac{1}{4}\, \bigg[ \frac{m_c^2}{u^2 M^4} \widetilde{\widetilde{\Theta}}(c(u,s_0)) + \frac{1}{u M^2}\,\widetilde{\Theta}(c(u,s_0))\bigg]\phi_{4;\phi}^\perp(u)
\,+\,\bigg[(4-2u)\,\bigg[\frac{\mathcal{C}}{u^3 M^4}\widetilde{\widetilde{\Theta}}(c(u,s_0))-\frac{2}{u^2 M^2}
\nonumber\\
&\hspace{0.42cm}\times\widetilde{\Theta}(c(u,s_0))\,\bigg]+2\bigg(\frac{2m_c^2}{u^2 M^4}\widetilde{\widetilde{\Theta}}(c(u,s_0))+\frac{1}{u M^2}\widetilde{\Theta}(c(u,s_0))\bigg)I_L(u)-\frac{2-u}{uM^2}\widetilde{\Theta}(c(u,s_0))H_3(u)\bigg\}
\nonumber\\
&\hspace{0.42cm}-\int\mathcal{D}\alpha_i\int_0^1dve^{\frac{m_{D_s^+}^2-s(X)}{M^2}}\frac{1}{2 X^2M^2}\,\Theta(c(X,s_0))\,\bigg[ \,(4v-1)\,\Psi_{4;\phi}^\perp(\underline{\alpha})\,-\,\widetilde{\Psi}_{4;\phi}^\perp(\underline{\alpha})
\,+\,4v\,\Phi_{3;\phi}^\perp(\underline{\alpha}\,) \bigg]\bigg\}.
\end{align}
\end{widetext}
where $\mathcal{H}=q^2/(m_{D_s^+}^2-m_\phi^2)$ and $\mathcal{C}=m_c^2+u^2m_\phi^2-q^2$. $s(\varrho)=[m_c^2-\bar{\varrho}(q^2-\varrho m_\phi^2)]/\varrho(\varrho=u,X)$ with $\bar{\varrho}=1-\varrho$, $X=a_1+a_3$. $c(\varrho,s_0)=\varrho s_0 - m_c^2 +\bar{\varrho}q^2-\varrho\bar{\varrho}m_\phi^2$. $\Theta(c(\varrho,s_0))$ is the usual step function, $\widetilde{\Theta}(c(\varrho,s_0))$ and $\widetilde{\widetilde{\Theta}}(c(\varrho,s_0))$ are defined via the integration
\begin{align}
\int_0^1&\frac{du}{u^2 M^2}e^{-s(u)/M^2}\widetilde{\Theta}(c(u,s_0))f(u)
\nonumber\\
=&\int_{u_0}^1\frac{du}{u^2 M^2}e^{-s(u)/M^2}f(u)+\delta(c(u_0,s_0)),
\\
\int_0^1&\frac{du}{2u^3M^4}e^{-s(u)/M^2}\widetilde{\widetilde{\Theta}}(c(u,s_0))f(u)
\nonumber\\
=&\int_{u_0}^1\frac{du}{2u^3M^4}e^{-s(u)/M^2}f(u)+\Delta(c(u_0,s_0)).
\end{align}
where the surface terms $\delta(c(u_0,s_0))$ and $\Delta(c(u_0,s_0))$ for the two-particle DAs are
\begin{align}
\delta(c(u_0,s_0))&=e^{-s_0/M^2}\frac{f_(u_0)}{\mathcal{C}_0},
\nonumber\\
\Delta(c(u_0,s_0))&=e^{-s_0/M^2}\bigg[ \frac{1}{2u_0M^2} \frac{f_{(u_0)}}{\mathcal{C}_0}
-\frac{u_0^2}{2\mathcal{C}_0}\frac{d}{du}\bigg( \frac{f(u)}{u\mathcal{C}}\bigg)\bigg|_{u=u_0}\bigg],
\end{align}
where $\mathcal{C}_0=m_c^2+u_0^2m_\phi^2-q^2$ and $u_0$ is the solution of $c(u_0,s_0)=0$ with $0\le u_0\le1$. There also exist surface terms from three-particle DAs, but their contributions are sufficiently negligible to be safely ignored. The simplified functions $I_L(u)$ and $H_3(u)$ are defined as follows:
\begin{align}
&I_L(u)= \int_0^udv\int_0^vdw\bigg[\phi_{3;\phi}^\parallel(w)-\frac{1}{2}\phi_{2;\phi}^\perp(w,\mu) -\frac{1}{2}\psi_{4;\phi}^\perp(w)\bigg],
\nonumber\\
&H_3(u)=\int_0^udv[\psi_{4;\phi}^\perp(v)-\phi_{2;\phi}^\perp(v,\mu)].
\end{align}
By utilizing these TFFs and considering the chirally suppressed lepton mass effects (which can be neglected), the differential decay width for $D_s^+\to \phi\ell^+\nu_\ell$ can be expressed as follows~\cite{Aliev:2004vf,Fu:2018yin,Zhong:2023cyc,Hu:2024tmc}:
\begin{align}
\frac{d\Gamma}{dq^2}
&=\frac{G_F^2 |V_{cs}|^2}{192\pi^3m_{D_s^+}^3} \lambda^{1/2} (m_{D_s^+}^2,m_\phi^2,q^2)q^2
\nonumber\\
&\hspace{0.38cm}\times[|H_+|^2 + |H_-|^2 + |H_0|^2]
\nonumber\\
&=\frac{d\Gamma_\mathrm{L}}{dq^2}+\frac{d\Gamma_\mathrm{T}^+}{dq^2}
+\frac{d\Gamma_\mathrm{T}^-}{dq^2}.
\end{align}
The $\phi$-meson has three polarization states: one longitudinal polarization and two transverse polarizations (right-handed and left-handed). These can be expressed as:
\begin{align}
\frac{d\Gamma_\mathrm{L}}{dq^2} &= \frac{G_F^2|V_{cs}|^2}{192\pi^2 m_{D_s^+}^3}\lambda^{1/2}(m_{D_s^+}^2,m_\phi^2,q^2)q^2\bigg|\frac{1}{2m_\phi\sqrt{q^2}}
\nonumber\\
&\hspace{0.4cm}\times \bigg[\,(m_{D_s^+}^2-m_\phi^2-q^2)\,(m_{D_s^+}^2+m_\phi)\,A_1(q^2)
\nonumber\\
&\hspace{0.4cm}-\frac{\lambda(m_{D_s^+}^2,m_\phi^2,q^2)}{m_{D_s^+}^2+m_\phi} \bigg]\bigg|^2,
\\
\frac{d\Gamma_\mathrm{T}^\pm}{dq^2}&= \frac{G_F^2|V_{cs}|^2}{192\pi^2 m_{D_s^+}^3}\lambda^{1/2}(m_{D_s^+}^2,m_\phi^2,q^2)q^2\bigg|(m_{D_s^+}+m_\phi)
\nonumber\\
&\hspace{0.4cm}\times A_1(q^2)\mp\frac{\lambda^{1/2}(m_{D_s^+}^2,m_\phi^2,q^2)}{m_{D_s^+}+m_\phi)}
V(q^2)\bigg|.
\end{align}
where $\lambda(x,y,z)=(x+y-z)^2-4xy$ is the K\"{a}llen function, and the combined transverse decay width is $\Gamma_\mathrm{T}=\Gamma_\mathrm{T}^+ + \Gamma_\mathrm{T}^-$.
After $q^2$ integratioin, we obtain the branching fraction for this process as
\begin{align}
\mathcal{B}(D_s^+\to \phi\ell^+\nu_\ell)= \tau_{D_s^+}\int_{m_\ell^2}^{(m_{D_s^+}-m_\phi)^2}\frac{d\Gamma}{dq^2}.
\end{align}
where $\tau_{{D_s^+}}$ is the $D_s^+$-meson lifetime.

In addition, the helicity amplitudes $H_\pm(q^2)$, $H_0(q^2)$ and $H_t(q^2)$ are defined as follows in terms of the TFFs:

\begin{align}
H_\pm(q^2)
&= \frac{\lambda^{1/2}(m_{D_s^+}^2,m_\phi^2,q^2)}{m_{D_s^+}+m_\phi}
\bigg[ \frac{(m_{D_s^+}+ m_\phi)^2}{ \lambda^{1/2}(m_{D_s^+}^2, m_\phi^2, q^2) }A_1(q^2)
\nonumber\\
&\hspace{0.4cm}\mp   V(q^2) \bigg],
\nonumber\\
H_0(q^2)
&=\frac{1}{2m_\phi\sqrt{q^2}}\bigg[ (m_{D_s^+}+m_\phi)(m_{D_s^+}^2\!\!-\!m_\phi^2\!-\!q^2)A_1(q^2)
\nonumber\\
&\hspace{0.4cm}-\frac{\lambda(m_{D_s^+}^2,m_\phi^2,q^2)}{m_{D_s^+}+m_\phi}A_2(q^2) \bigg],
\nonumber\\
H_t(q^2)
&=\frac{\lambda^{1/2}(m_{D_s^+}^2,m_\phi^2,q^2)}{\sqrt{q^2}}A_0(q^2).
\end{align}

Based on the helicity amplitudes defined above, the forward-backward asymmetry $A_{FB}^\ell(q^2)$, the lepton-side convexity parameter $C_F^\ell(q^2)$, and the longitudinal (transverse) polarization of the final charged lepton $P_{L(T)}^\ell(q^2)$, as well as the longitudinal (transverse) polarization fraction of the final $\phi$-meson $F_{L(T)}^\ell(q^2)$ in the semileptonic decay $D_s^+\to\phi\ell^+\nu_\ell$, can be expressed as follows, according to Refs.~\cite{Ivanov:2019nqd, Faustov:2019mqr}
\begin{align}
A_{\mathrm{FB}}^\ell(q^2)
&=\frac{\int_0^1 \cos\theta d\Gamma/ d\cos\theta-\int_{-1}^0 d\cos\theta d\Gamma/d\cos\theta}{\int_0^1d\cos\theta d\Gamma/d\cos\theta+\int_{-1}^0d\cos\theta d\Gamma/d\cos\theta}
\nonumber\\
&=\frac{3}{4}\frac{\mathcal{H}_P-2\delta\mathcal{H}_{\mathrm{SL}}}{\mathcal{H}_
{\mathrm{total}}},
\\
C_\mathrm{F}^\ell(q^2)
&=\frac{3}{4}(1-\delta)\frac{\mathcal{H}_\mathrm{U}-2\mathcal{H}_\mathrm{L}}
{\mathcal{H}_\mathrm{total}},
\\
P_\mathrm{L}^\ell(q^2)
&=\frac{(\mathcal{H}_\mathrm{U}+\mathcal{H}_\mathrm{L})(1-\frac{\delta}{2})
-\frac{3\delta}{2}\mathcal{H}_\mathrm{S}}{\mathcal{H}_\mathrm{total}},
\\
P_\mathrm{T}^\ell(q^2)
&=-\frac{3\pi m_\ell}{8\sqrt{q^2}}\frac{\mathcal{H}_\mathrm{P}+2\mathcal{H}_\mathrm{SL}}
{\mathcal{H}_\mathrm{total}},
\\
F_\mathrm{L}^\ell(q^2)
&=\frac{\mathcal{H}_\mathrm{L}(1+\frac{\delta}{2})+\frac{3\delta}{2}\mathcal{H}_\mathrm{S}}
{\mathcal{H}_\mathrm{total}}.
\end{align}
where the $m_\ell$ with $\ell=(e,\mu)$ is the lepton mass and the helicity structure functions $\mathcal{H}_i$ are defined as
\begin{align}
&\mathcal{H}_\mathrm{U} =|H_+|^2+|H_-|^2,
\nonumber\\
&\mathcal{H}_\mathrm{L} =|H_0|^2,
\nonumber\\
&\mathcal{H}_\mathrm{P} =|H_+|^2-|H_-|^2,
\nonumber\\
&\mathcal{H}_\mathrm{SL} =\mathrm{Re}(H_0H_t^\dagger),
\nonumber\\
&\mathcal{H}_\mathrm{S}=|H_t|^2.
\end{align}
In addition, the total helicity amplitude takes the
\begin{align}
\mathcal{H}_\mathrm{total}= (\mathcal{H}_\mathrm{U}+\mathcal{H}_\mathrm{L})(1+\delta/2)+3\delta\mathcal{H}_\mathrm{S}/2
\end{align}
with $\delta=m_\ell^2/q^2$ and $F_\mathrm{T}^\ell(q^2)=1-F_\mathrm{L}^\ell(q^2)$.

\section{Numerical Analysis}\label{sec:analysis}
Before performing the numerical analysis, we adopt the following parameters. We take the $\phi $-meson decay constant $f_{\phi}^{\parallel} = 215 \pm 5 ~\mathrm{MeV}$~\cite{Ball:2007zt}, According to the Particle Data Group (PDG) ~\cite{ParticleDataGroup:2024cfk}, we take the $\phi$-meson mass $m_{\phi} = 1019.461 \pm 0.016~\mathrm{MeV}$, the charm-quark current mass $m_c(\bar{m}_c) = 1.2730 \pm 0.0046~\mathrm{GeV}$, the $s$-quark current mass $m_s(2~\mathrm{GeV}) = 93.5 \pm 0.8~\mathrm{MeV}$, the $D_s$-meson mass $m_{D_s} = 1968.35 \pm 0.07~\mathrm{MeV}$ and the $D_s$-meson decay constant $f_{D_s} = 249.9 \pm 0.5~\mathrm{GeV}$. The factorization scale $\mu $ is set as the typical momentum transfer of $D_s \to \phi $, {\it i.e.} $\mu_k \approx (m_{D_s^+}^2 - \bar{m_c}^2)^{1/2} \approx 1.5~\mathrm{GeV}$.
In addition, the renormalization group equation (RGE) of the Gegenbauer moments of the $\phi_{2;\phi}^{\bot}(x,\mu_0)$ leading-twist LCDA is~\cite{Ball:2004ye}:
\begin{align}
a_n^{2;\phi}(\mu) = a_n^{2;\phi}(\mu_0) \left[ \frac{\alpha_s(\mu)}{\alpha_s(\mu_0)} \right]^{\frac{\gamma_n}{2 \beta_0}},
\end{align}
with
\begin{align}
\gamma_n &= C_F \left( 1-\frac{2}{(n+1)(n+2)}+4\sum_{m=2}^{n+1} \frac{1}{m} \right)\nonumber\\
\beta_0 &= 11 - 2/3 n_f
\end{align}
where $n_f$ is the number of flavours involved. The dominant $\phi$-meson transverse leading-twist LCDA $\phi_{2;\phi}^{\bot}(x,\mu_0)$ can be derived from its light-cone wavefunction (LCWF), which is related via the relation
\begin{align}
\phi_{2;\phi}^{\bot} (x,\mu_0) =
\frac{2 \sqrt{3}}{\tilde{f}_{\phi}^{\bot}} \int_{|{\bf k}_{\bot}|^2 \le \mu_0^2}
\frac{d {\bf k}_{\bot}}{16 \pi^3} \psi_{2;\phi}^{\bot}(x,{\bf k}_{\bot}),
\end{align}
where $\tilde{f}_{\phi}^{\bot} = f_{\phi}^{\bot}/ \mathcal{C}_{\phi}^{\bot}$ is the reduced vector decay constant with $\mathcal{C}_{\phi}^{\bot} = \sqrt{3}$.
In the LCHO model~\cite{Huang:2013yya}, one can divide the $\phi$-meson LCWF into two parts, {\it i.e.} the redial part and spin-space part accordingly. Based on the BHL prescription~\cite{Brodsky:1981jv}, the LCHO model of the $\phi$ meson leading-twist WF can be given~\cite{Wu:2010zc,Wu:2011gf}:
\begin{align}
\psi_{2;\phi}^{\bot}(x,{\bf k}_{\bot}) = \sum_{h_1 h_2} \chi_{\phi}^{h_1 h_2} ({\bf k}_{\bot}) \psi_{2;\phi}^R (x,{\bf k}_{\bot}),
\end{align}
where $\chi_{\phi}^{h_1 h_2} ({\bf k}_{\bot})$ stands for the spin-space WF, $h_1$ and $h_2$ are the helicity states of the two constitute quarks in $\phi$. The  $\chi_{\phi}^{h_1 h_2} ({\bf k}_{\bot})$ comes from the Wigner-Melosh rotation~\cite{Huang:1994dy,Huang:2004su,Wu:2005kq}. $\psi_{2;\phi}^R (x,\mathbf{k}_{\bot})$ indicates the spatial WF, which can be divided into a $\mathbf{k}_{\bot}$-dependent part and a $x$-dependent part. Therefore, the spin-space WF and the spatial WF can be expressed as respectively:
\begin{align}
&\chi_{\phi}^{h_1 h_2} ({\bf k}_{\bot}) = \frac{\hat{m}_s}{\sqrt{{\bf k}_{\bot}^2 + \hat{m}_s^2}},
\\
&\psi_{2;\phi}^R (x,{\bf k}_{\bot}) \propto \left[ 1+ B_{2;\phi}^{\bot} C_2^{3/2}(\xi) \right] \nonumber\\
&\qquad\qquad~ \times \exp \left[ -b_{2;\phi}^{\bot 2} \left( \frac{\mathbf{k}_{\bot}^2 + \hat{m}_s^2}{x} + \frac{{\bf k}_{\bot}^2+\hat{m}_q^2}{\bar{x}} \right) \right].
\end{align}
where $C_2^{3/2}(\xi)$ is Gegenbauer polynomial, $\xi = 2x-1$, $x$ stands for the $s$-quark momentum fraction of the meson and $\bar{x} = 1-x$ stands for that of the light-quark $q$. $\hat{m}_s$ is the $s$-quark mass and $\hat{m}_q$ is the light-quark mass. Further, one can obtain the LCDA
\begin{align}
\hspace{-0.2cm}\phi_{2;\phi}^{\bot} (x,\mu_0) &= \frac{3 A_{2;\phi}^{\bot} \hat{m}_s \sqrt{x \bar{x}} }{4 \pi^{3/2} f_{\phi}^{\bot} b_{2;\phi}^{\bot} } \,\left[ 1\,+\,B_{2;\phi}^{\bot} C_2^{3/2}(\xi)\, \right]
\nonumber\\
&\times \!\left[\! \mathrm{Erf}\!\left( b_{2;\phi}^{\bot} \sqrt{\frac{\mu_0^2+\hat{m}_s^2}{x\bar{x}}}\right)\!-\!\mathrm{Erf}\! \left(b_{2;\phi}^{\bot} \frac{\hat{m}_s}{\sqrt{x\bar{x}}} \right) \!\right],
\end{align}
where $\mathrm{Erf} = \frac{2}{\sqrt{\pi}} \int_0^x e^{-t^2}dt$ and the constituent quark mass $\hat{m}_s ~\mathrm{or}~ \hat{m}_q \simeq 370 ~\mathrm{MeV}$. In addition to the normalization condition, the average value of the squared transverse momentum can be regarded as another constraint, which is defined as
\begin{align}
\langle {\bf k}^2 \rangle_{2;\phi}
= \frac{\int dxd^2{\bf k}_{\bot}|{\bf k}_{\bot}|^2 |\psi_{2;\phi}^{\bot}(x,{\bf k}_{\bot})|^2 }{\int dxd^2{\bf k}_{\bot} |\psi_{2;\phi}^{\bot}(x,{\bf k}_{\bot})|^2 }.
\end{align}
Here, the value of $\langle {\bf k}_{\bot}^2 \rangle_{2;\phi} \sim 0.37~\mathrm{GeV}^2$~\cite{Wu:2010zc}. In general, the light meson's transverse leading-twist LCDA can be expanded into a series of Gegenbauer polynomials. The Gegenbauer moments $a_n^{\bot}$ at the initial scale $\mu_0$ have been obtained by the following formula
\begin{align}
a_n^{\bot} (\mu_0)
= \frac{\int_0^1 dx \phi_{2;\phi}^{\bot} (x,\mu_0)C_n^{3/2}(\xi)}
{\int_0^1 dx 6x\bar{x}[C_n^{3/2}(\xi)]^2},
\end{align}
where $\mu_0\sim 1~\mathrm{GeV}$ stands for some initial scale.
Besides, the normalization of the wave function is as follows
\begin{align}
\int_0^1 dx\int \frac{d^2 {\bf k}_{\bot}}{16\pi^3}\psi^{\bot}_{2;\phi}(x,{\bf k}_\bot) = \frac{f_\phi^\bot}{2\sqrt{6}}.
\end{align}

\begin{figure}[t]
\begin{center}
\includegraphics[width=0.40\textwidth]{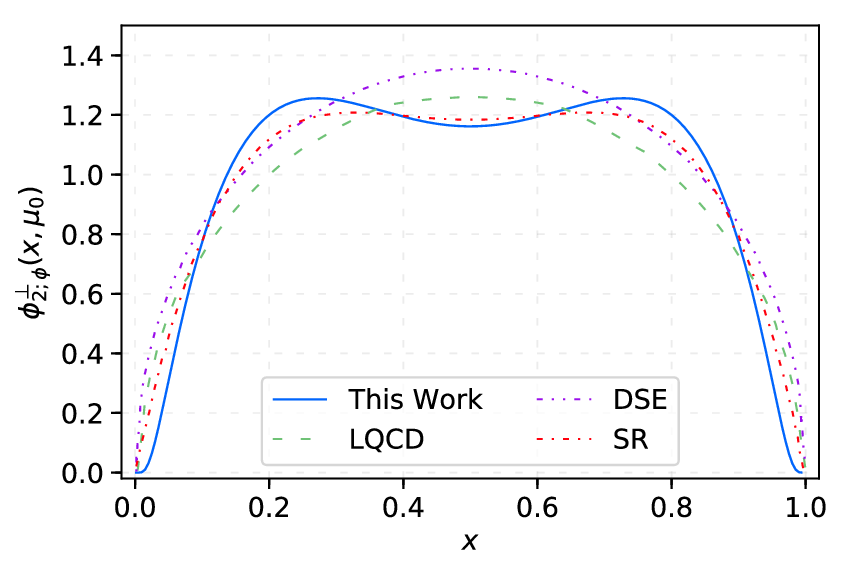}
\end{center}
\caption{The transverse leading-twist LCDA $\phi_{2;\phi}^{\bot}(x,\mu_0)$ at the scale $\mu_0=1.5~\mathrm{GeV}$. As a comparison, the results derived from the SR~\cite{Ball:2007zt}, the DSE~\cite{Gao:2014bca} and LQCD~\cite{Hua:2020gnw} approaches have been presented.}
\label{Fig:DA}
\end{figure}

The LCHO model has three undetermined parameters $A_{2;\phi}^{\bot}$, $b_{2;\phi}^{\bot}$ and $B_{2;\phi}^{\bot}$. In addition to the normalization condition and the squared transverse momentum $\langle {\bf k}_\bot^2 \rangle_{2;\phi} $, one shall adopt the values of the second Gegenbauer moment $a_2^\bot (1~\mathrm{GeV})=0.14(7)$~\cite{Ball:2007zt}. Based on this, the behavior of the dominant $\phi$-meson transverse leading-twist LCDA $\phi_{2;\phi}^{\bot}(x,\mu_0)$ can be determined and showed in Fig.~\ref{Fig:DA}. The corresponding parameter is $A_{2;\phi}^{\bot}=14.1835$, $B_{2;\phi}^{\bot}=0.163502$ and $b_{2;\phi}^{\bot}=0.625318$, respectively.
The parameter $B_{2;\phi}^{\bot}$ quantitatively controls the behavior of the LCDA. A larger $B_{2;\phi}^{\bot}$ value manifests a double-peak structure, while a smaller $B_{2;\phi}^{\bot}$ exhibits single-peak behavior.
The Fig.~\ref{Fig:DA} also includes comparisons with the SR~\cite{Ball:2007zt}, the Dyson-Schwinger equations (DSE)~\cite{Gao:2014bca} and LQCD~\cite{Hua:2020gnw}. Notably, our results exhibit a double-peak behavior consistent with the SR predictions.
Further, we adopt the LCDA for discussion in the following.

As two important input parameters , the continuum threshold parameter $s_0$ and the the Borel parameter $M^2$, in QCD sum rules~\cite{Tian:2024ubt,Wang:2024oty,Zhong:2023cyc,Hu:2023pdl,Tian:2023vbh}, we adopt the following criteria to obtain reasonable input parameters:
\begin{itemize}
\item The continuum contribution is less than $35\%$;
\item The contribution of the high-twist LCDAs is no more than $5\%$;
\item The value of TFFs is stable in the Borel window;
\end{itemize}

\begin{table}[t]
\footnotesize
\begin{center}
\caption{Comparison of various theoretical for the $D_s^+ \to \phi$ TFFs $A_1(q^2)$, $A_2(q^2)$ and $V(q^2)$ at the large recoil point $q^2 = 0$. }
\label{Tab:TFFs}
\begin{tabular}{l l l l l}
\hline
\hline
~~~~~~~~~~~~~~~~~~~~~~~~~~~~~~~~~~&$A_1(0)$~~~~~~~~~~~~~~~~~&$A_2(0)$~~~~~~~~~~~~~~~~~&$V(0)$          \\
\hline
This Work                                      &$0.600_{-0.020}^{+0.022}$          &$0.567_{-0.056}^{+0.050}$                                               &$0.910_{-0.024}^{+0.024}$
\\
HQEFT'06(1)~\cite{Wu:2006rd}                                     &$0.569_{-0.049}^{+0.046}$          &$0.304_{-0.017}^{+0.021}$                                               &$0.778_{-0.062}^{+0.057}$
\\
HQEFT'06(2)~\cite{Wu:2006rd}                                     &$0.605_{-0.042}^{+0.043}$          &$0.583_{-0.036}^{+0.038}$                                               &$0.951_{-0.053}^{+0.053}$
\\
LCSR'04~\cite{Aliev:2004vf}                    &$0.54(9)$          &$0.57(9)$      &$0.70(10)$
\\
LCSR'24(1)~\cite{Hu:2024tmc}           &$0.514_{-0.016}^{+0.024}$   &$0.438_{-0.080}^{+0.093}$  &$0.902_{-0.024}^{+0.040}$
\\
LCSR'24(2)~\cite{Hu:2024tmc}           &$0.512_{-0.020}^{+0.030}$   &$0.402_{-0.067}^{+0.078}$  &$0.882_{-0.036}^{+0.040}$
\\
HM$\chi$T'05~\cite{Fajfer:2005ug}                   &$0.61$          &$0.32$          &$1.10$
\\
CQM'00~\cite{Melikhov:2000yu}                                    &$0.64$          &$0.47$                                               &$1.10$
\\
3PSR'03~\cite{Du:2003ja}                                    &$0.55(15)$          &$0.59(11)$                                               &$1.21(33)$
\\
CLFQM'11~\cite{Verma:2011yw}                                  &$0.69$          &$0.57$                                               &$0.98$
\\
LFQM'19~\cite{Chang:2019mmh}                                     &$0.77$          &$0.66$
&$1.24$
\\
CCQM'19~\cite{Ivanov:2019nqd}                    &$0.68$         &$0.67$           &$0.91$
\\
RQM'20~\cite{Faustov:2019mqr}           &$0.643$        &$0.492$        &$0.999$
\\
LQCD'01~\cite{Gill:2001jp}              &$0.63(2)$          &$0.62(5)$        &$0.85(4)$
\\
LQCD'11~\cite{Donald:2011ff}             &$0.594(22)$        &$0.401(80)$     &$0.903(67)$
\\
LQCD'13~\cite{Donald:2013pea}                     &$0.615(24)$  &$0.457(78)$   &$1.059(124)$
\\
\hline
\hline
\end{tabular}
\end{center}
\end{table}

Based on the criteria for determining $s_0$ and $M^2$ in the LCSR approach, one takes the $s_0^{A_1}=9.5\pm 0.5~\mathrm{GeV^2}$, $s_0^{A_2} = 4.4\pm 0.25~\mathrm{GeV^2}$, $s_0^{V}=7.6\pm 0.5~\mathrm{GeV^2}$ and $M_{A_1}^2=4.0\pm0.8~\mathrm{GeV^2}$, $M_{A_2}^2=2.9\pm0.2~\mathrm{GeV^2}$, $M_{V}^2=3.7\pm0.5~\mathrm{GeV^2}$.
Then, the TFFs at the large recoil point $q^2=0$, {\it i.e.} $A_{1,2}(0)$ and $V(0)$ are present in Table~\ref{Tab:TFFs}.
These values at the large recoil point are crucial for connecting theoretical models with experimental measurements. The differences in describing hadronic internal structures among various theoretical approaches are directly reflected at the large recoil point.
Our calculated TFFs exhibit the large recoil point uncertainties ranging from $2.6\%$ to $8.9\%$. The uncertainties are coming from the squared average for all the input parameters.
In Table~\ref{Tab:TFFs}, the results given by several groups have also been presented {\it i.e.} HQEFT'06~\cite{Wu:2006rd}, LCSR~\cite{Aliev:2004vf,Hu:2024tmc}, HM$\chi$T~\cite{Fajfer:2005ug}, CQM~\cite{Melikhov:2000yu}, 3PSR~\cite{Du:2003ja}, CLFQM~\cite{Verma:2011yw}, LFQM~\cite{Chang:2019mmh}, CCQM~\cite{Ivanov:2019nqd}, RQM~\cite{Faustov:2019mqr}, LQCD~\cite{Gill:2001jp, Donald:2011ff,Donald:2013pea}.
The comparison about the every results from Table~\ref{Tab:TFFs} indicate that the TFFs of our predictions is consistent with many approaches within errors.
Our results for $A_1(0)$ and $V(0)$ exhibit good agreement with the LQCD'11~\cite{Donald:2011ff} predictions. Moreover, the $A_2(0)$ values show reasonable consistency with both the LCSR'04~\cite{Aliev:2004vf} and CLFQM'11~\cite{Verma:2011yw} results.
This observation confirms that our results demonstrate reasonable validity for the value at the large recoil point.
Regarding the $\omega-\phi$ mixing effects, we adopt a mixing angle of $\phi_V=3.4^\circ$ for the present calculation.
Our analysis reveals that the mixing angle $\varphi_V$ contributes $0.044014\%$ to $A_1(0)$, $0.044091\%$ to $A_2(0)$ and $0.044066\%$ to $V(0)$.
The negligible magnitude of its contribution justifies our neglect of the mixing angle effects.
Therefore, we adopt the ideal pure-state approximation treating the $\phi$ meson as a pure $s\bar{s}$ state in subsequent calculations.
We anticipate that near-future breakthrough in understanding the $\phi$ meson's properties will significantly advance our comprehension of $\phi$-meson-related decays.

\begin{table}[t]
\footnotesize
\begin{center}
\caption{The prediction for the ratio $\gamma_V$ and $\gamma_2$ of $D_s^+\to\phi$ decay within uncertain. Meanwhile, both theoretical and experimental results are presented for comparison.}
\label{Tab:Ratio}
\begin{tabular}{l l l l l}
\hline
\hline
~~~~~~~~~~~~~~~~~~~~~~~~~~~~~~~~~~&$\gamma_V$~~~~~~~~~~~~~~~~~~~~~~~~~~~~~~~~~~&$\gamma_2$          \\
\hline
This Work               &$1.517_{-0.015}^{+0.011}$          &$0.945_{-0.064}^{+0.047}$                                               \\
HQEFT'06(1)~\cite{Wu:2006rd}       &$1.37_{-0.21}^{+0.24}$      &$0.53_{-0.06}^{+0.10}$
\\
HQEFT'06(2)~\cite{Wu:2006rd}       &$1.57_{-0.18}^{+0.21}$      &$0.96_{-0.12}^{+0.14}$
\\
LCSR'24(1)~\cite{Hu:2024tmc}      &$1.755_{-0.005}^{+0.008}$   &$0.852_{-0.133}^{+0.135}$
\\
LCSR'24(2)~\cite{Hu:2024tmc}           &$1.723_{-0.021}^{+0.023}$   &$0.785_{-0.104}^{+0.100}$
\\
HM$\chi$T'05~\cite{Fajfer:2005ug}        &$1.8$          &$0.52$
\\
3PSR'03~\cite{Du:2003ja}               &$2.20(85)$           &$1.07(0.43)$
\\
CCQM'19~\cite{Ivanov:2019nqd}           &$1.34(27)$         &$0.99(20)$
\\
RQM'20~\cite{Faustov:2019mqr}           &$1.56$             &$0.77$
\\
FOCUS'04~\cite{FOCUS:2004gfa}           &$1.549(250)(145)$  &$0.713(202)(266)$
\\
BaBar'06~\cite{BaBar:2006typ}          &$1.636(67)(38)$     &$0.705(56)(29)$
\\
BaBar'08~\cite{BaBar:2008gpr}          &$1.807(46)(65)$     &$0.816(36)(30)$
\\
BESIII'23~\cite{BESIII:2023opt}        &$1.58(17)(2)$        &$0.71(14)(2)$
\\
LQCD'01~\cite{Gill:2001jp}              &$1.35(7)$          &$0.98(8)$
\\
LQCD'11~\cite{Donald:2011ff}             &$1.52(12)$        &$0.62(12)$
\\
LQCD'13~\cite{Donald:2013pea}            &$1.72(21)$        &$0.74(12)$
\\
\hline
\hline
\end{tabular}
\end{center}
\end{table}
Furthermore, we provide two particularly interesting ratios $\gamma_V=V(0)/A_1(0)$ and $\gamma_2 = A_2(0)/A_1(0)$ in Table~\ref{Tab:Ratio}.
Meanwhile, a comparative analysis is conducted between theoretical predictions (HQEFT~\cite{Wu:2006rd}, LCSR~\cite{Aliev:2004vf,Hu:2024tmc}, 3PSR~\cite{Du:2003ja}, CCQM~\cite{Ivanov:2019nqd}, RQM~\cite{Faustov:2019mqr}, LQCD'01~\cite{Gill:2001jp,Donald:2011ff,Donald:2013pea}) and experimental results (from the FOCUS Collaboration~\cite{FOCUS:2004gfa}, the BaBar Collaboration~\cite{BaBar:2006typ,BaBar:2008gpr} and the BESIII Collaboration~\cite{BESIII:2023opt}).
For $\gamma_V$, our results show good consistency with the FOCUS'04~\cite{FOCUS:2004gfa} and BESIII'23~\cite{BESIII:2023opt} measurements. Regarding theoretical comparisons, we agree well with the LQCD'11~\cite{Donald:2011ff} predictions.
Moreover, for $\gamma_2$, our results exhibit good agreement with both the HQEFT'06~\cite{Wu:2006rd} and LQCD'01~\cite{Gill:2001jp} calculations.
These ratios effectively eliminate the common uncertainties from the overall normalization of TFFs, thereby more sensitively revealing the differences in hadronic internal dynamics descriptions among various models.
Meanwhile, the experimentally measured ratios provide high-precision benchmarks for validating theoretical models.

\begin{figure}[t]
\begin{center}
\includegraphics[width=0.40\textwidth]{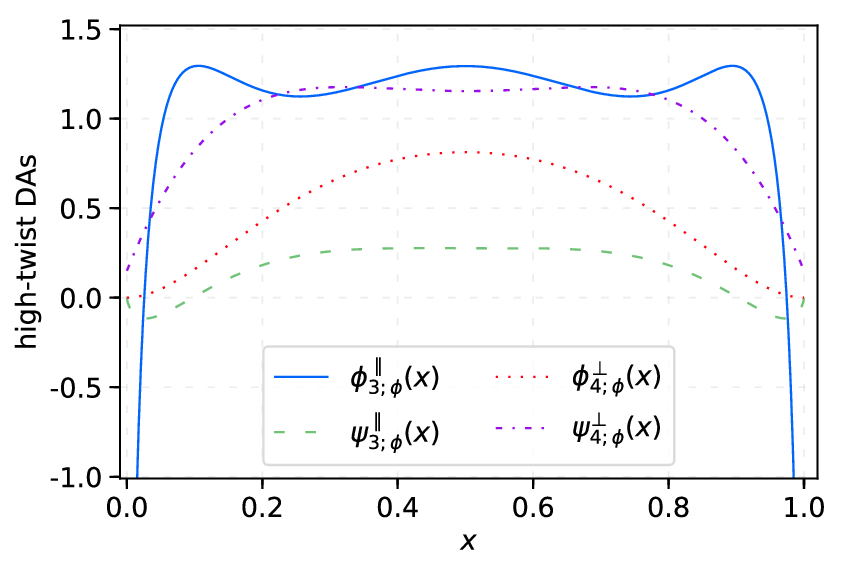}
\end{center}
\caption{The chiral-odd high-twist DAs of $\phi$-meson at the scale $\mu_0=1~\mathrm{GeV}$.  }
\label{Fig:twist34}
\end{figure}

\begin{table}[h]
\footnotesize
\begin{center}
\caption{Different twist contributions for the $D_s^+\to\phi\ell^+\nu_\ell$ TFFs.}
\label{Tab:DA-contribution}
\begin{tabular}{p{1.5cm} p{1.5cm} p{1.5cm} p{1.5cm} p{1.5cm}}
\hline
\hline
&$A_1(0)$ &$A_2(0)$ &$V(0)$          \\
\hline
Twist-2              &$0.558$          &$0.923$          &$1.170$
\\
Twist-3              &$0.027$          &$-0.098$         &$-$
\\
Twist-4              &$0.013$          &$-0.352$         &$-0.260$
\\
$I_L(x)$                &$0.003$          &$0.091$          &$-$
\\
$H_3(x)$                &$-0.001$          &$0.002$         &$-$
\\
Total                &$0.600$          &$0.567$          &$0.910$
\\
\hline
\hline
\end{tabular}
\end{center}
\end{table}
For the TFFs derived from the right-handed chiral correlator, the relative importance of different twist DAs follows these key trend: twist-2 $\gg$ twist-3 $\sim$ twist-4.
The twist-3 and twist-4 expressions employed in this work are derived from Refs.~\cite{Ball:2007rt,Ball:2007zt}.
Correspondingly, we present the behaviors of twist-3 and twist-4 in the Fig.~\ref{Fig:twist34}.
As shown, the longitudinal distribution amplitude $\phi_{3;\phi}^\parallel(x)$ exhibits strong suppression and approaches negative values near the endpoints ($x \to 0,1$).
This behavior introduces significant suppression when integrated over the entire domain.
Similarly, the DA $\psi_{3;\phi}^\parallel(x)$ does not vanish completely near the endpoints but exhibits a small negative-value region, resulting in minor suppression upon integration over the entire domain.
For the transverse twist-4 DAs $\psi_{4;\phi}^\perp(x)$ and $\phi_{4;\phi}^\perp(x)$, a pronounced peak emerges with significantly enhanced magnitude.
In particular, $\psi_{4;\phi}^\perp(x)$ exhibits a double-peak structure with both higher peak magnitudes and broader high-value regions.
This characteristic behavior leads to significant enhancement when integrated over the entire domain.
Furthermore, one presents the contributions of different twist LCDAs to the $D_s^+\to\phi$ TFFs in Table~\ref{Tab:DA-contribution}.
For both $A_1(0)$ and $V(0)$, the contributions from different twist DAs follow this key trend: twist-2 $\gg$ twist-3 $\sim$ twist-4.
For $A_2(0)$, we find that the twist-4 DA contributes more significantly than the twist-3 DA component.
Our systematic analysis demonstrates that this phenomenon arises because surface terms account for 61.2\% of the total contributions in the twist-4 calculation procedure.
For the two simplified distribution functions $I_L(x)$ and $H_3(x)$, their contributions are significantly smaller compared to those from the twist-2.
The dominance of the twist-2 term indicates a more convergent twist-expansion could be achieved by using the chiral correlator.

\begin{figure}[t]
\begin{center}
\includegraphics[width=0.40\textwidth]{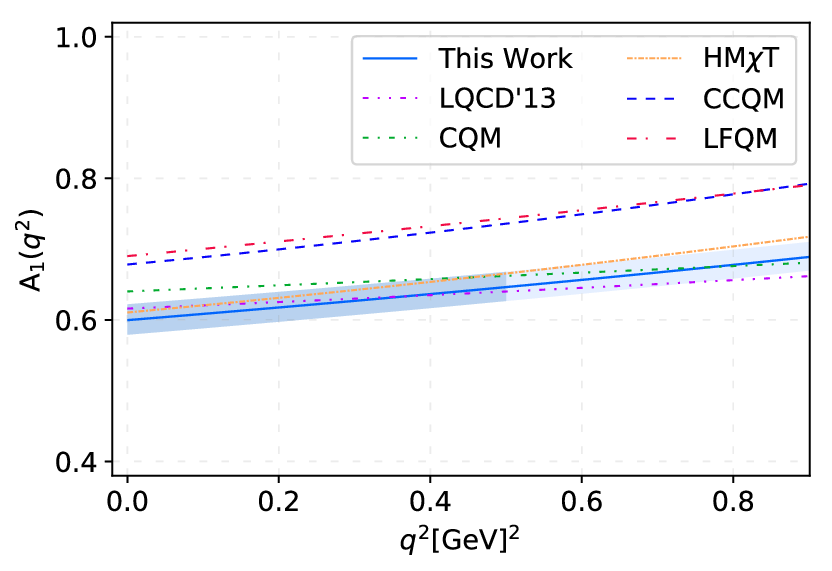}
\includegraphics[width=0.40\textwidth]{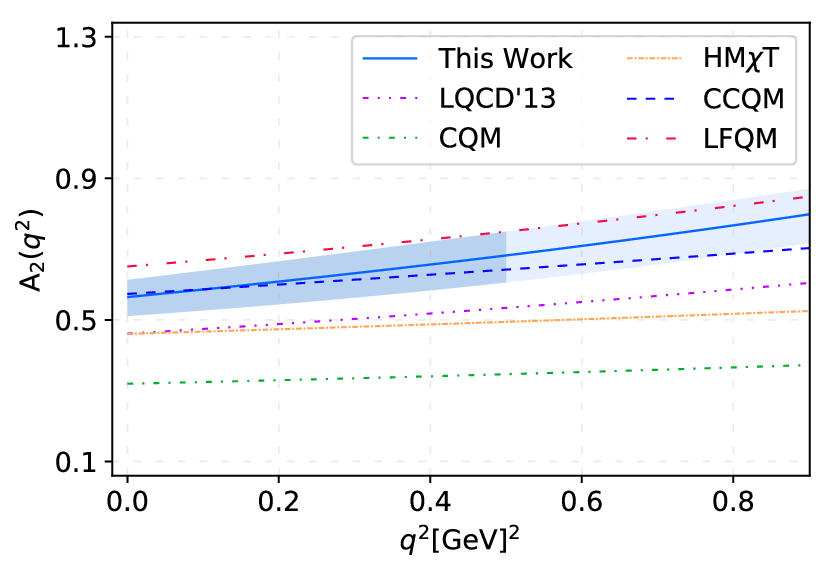}
\includegraphics[width=0.40\textwidth]{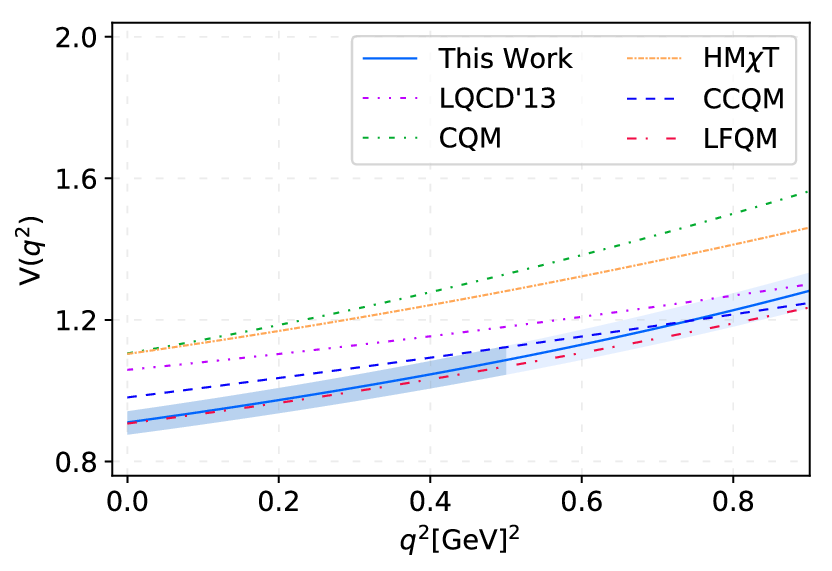}
\end{center}
\caption{The extrapolated TFFs $A_1(q^2)$, $A_2(q^2)$ and $V(q^2)$ in the whole $q^2$-region, where the solid line are central values the shaded bands are corresponding uncertainties. In addition, it is compared with other predictions. }
\label{Fig:TFFs}
\end{figure}

Comparative studies of TFFs across the entire $q^2$-region can reveal the consistency of theoretical approaches in the transition region between non-perturbative and perturbative regimes.
For the extrapolation of TFFs, we employ the SSE parameterizations. Different parameterizations schemes can affect the extracted slope.
In Fig.~\ref{Fig:TFFs}, we present the extrapolated behavior of TFFs across the entire $q^2$ region, which also includes the HM$\chi$T~\cite{Fajfer:2005ug}, the CQM~\cite{Melikhov:2000yu}, the LFQM~\cite{Chang:2019mmh}, the CCQM~\cite{Ivanov:2019nqd} and the LQCD'13~\cite{Donald:2013pea}.
The shaded bands in our prediction arises from input parameter variations, while other groups show their central predictions.
The slope of TFFs in Fig.~\ref{Fig:TFFs} exhibits a slightly steeper behavior compared to other theoretical predictions.
The advantage of employing the SSE parameterizations lies in its ability to effectively translate the near-threshold behavior of TFFs into constraints on the expansion coefficients.
The TFF $A_1(q^2)$ shows good agreement with both CQM~\cite{Melikhov:2000yu} and LQCD'13~\cite{Donald:2013pea} within uncertainties, while $A_2(q^2)$ and $V(q^2)$ exhibit behavior largely consistent with LFQM~\cite{Chang:2019mmh} prediction.

\begin{figure}[t]
\begin{center}
\includegraphics[width=0.40\textwidth]{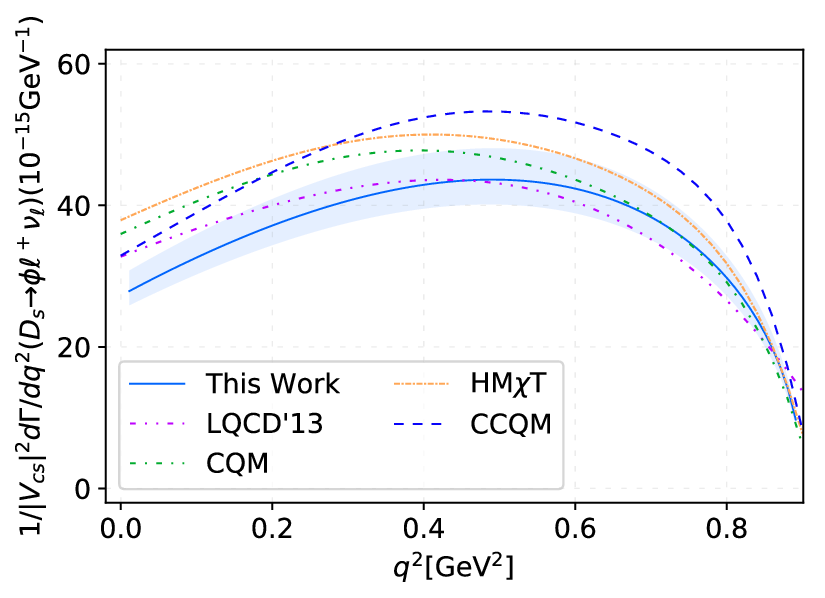}
\end{center}
\caption{The differential decay width $1/|V_{cs}|^2 d\Gamma(D_+^s \to \phi\ell^+\nu_\ell)/dq^2$ as a function of $q^2$, where the solid lines are central values and the shaded bands are corresponding uncertainties. In addition, it is compared with other predictions. }
\label{Fig:dGamma}
\end{figure}

Through those TFFs, we calculate the differential decay width $1/|V_{cs}|^2 d\Gamma(D_+^s \to \phi\ell^+\nu_\ell)/dq^2$, and the results are presented in Fig.~\ref{Fig:dGamma}. The HM$\chi$T~\cite{Fajfer:2005ug}, the LQCD'13~\cite{Donald:2013pea}, the CCQM~\cite{Ivanov:2019nqd} and the CQM~\cite{Melikhov:2000yu} predictions are also presented.
The results indicate that in the high $q^2$-region, our differential decay width shows good agreement with other results. Although there is a noticeable deviation in the low $q^2$-region, this is reasonable. In the high $q^2$-region, our behavioral prediction shows good agreement with the LQCD'13~\cite{Donald:2013pea} within uncertainty bounds, while also being largely consistent with both the CQM~\cite{Melikhov:2000yu} and the HM$\chi$T~\cite{Fajfer:2005ug} results.
\begin{figure*}[t]
\begin{center}
\includegraphics[width=0.40\textwidth]{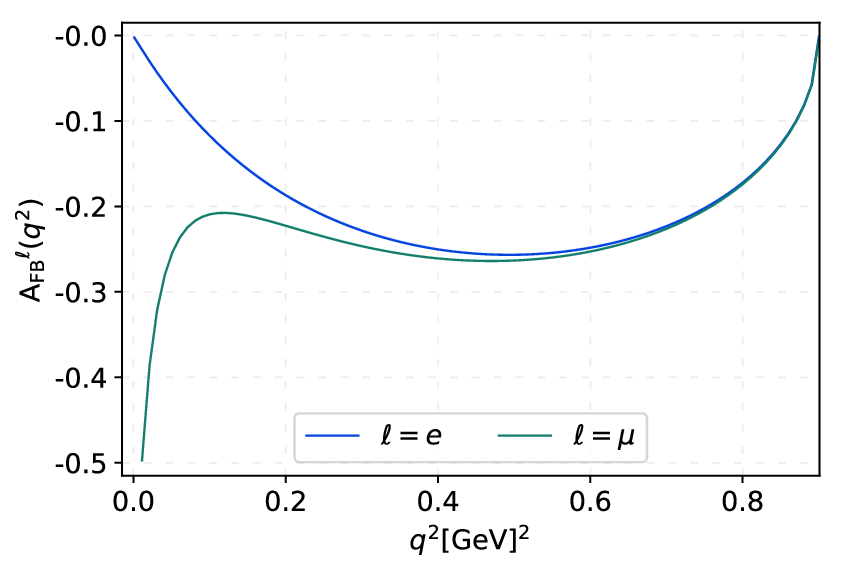}
\includegraphics[width=0.40\textwidth]{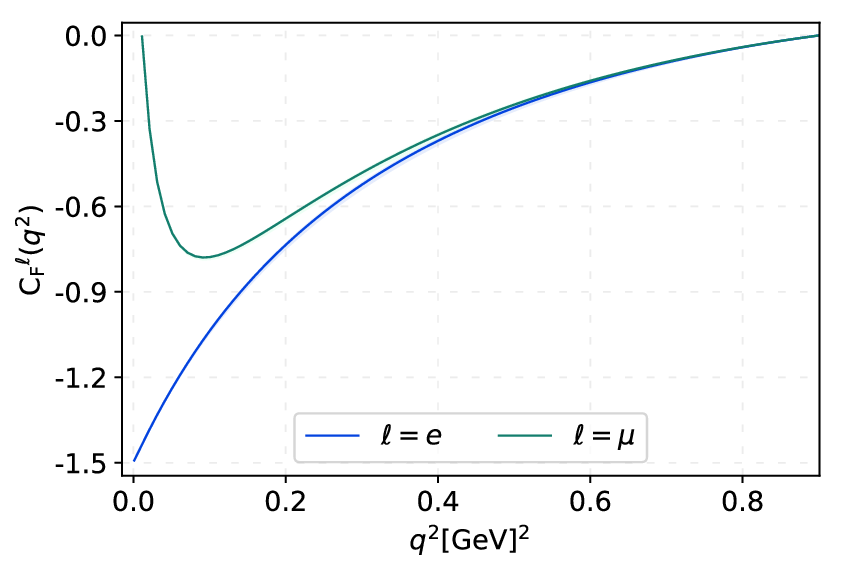}
\includegraphics[width=0.40\textwidth]{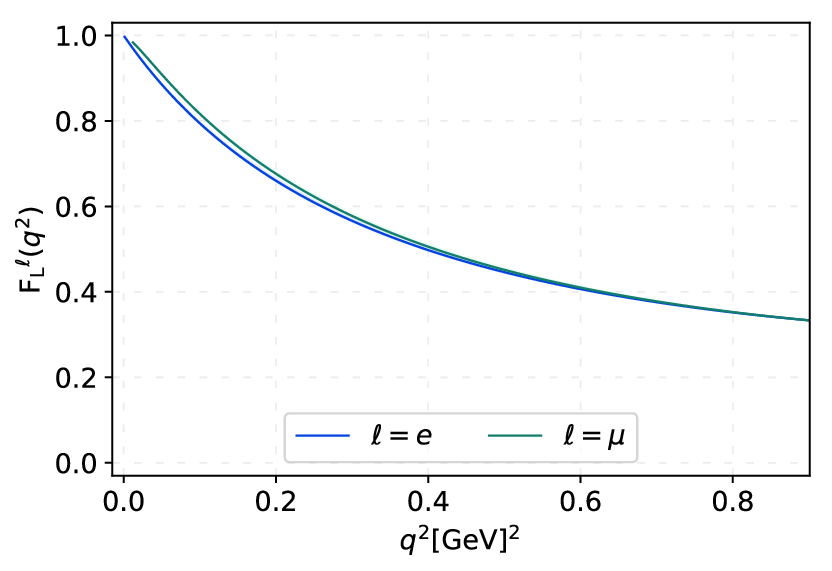}
\includegraphics[width=0.40\textwidth]{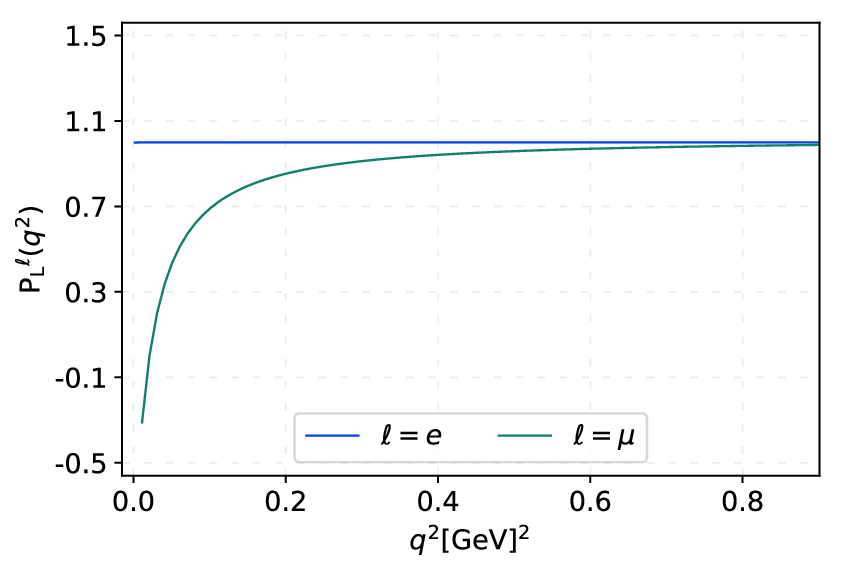}
\includegraphics[width=0.40\textwidth]{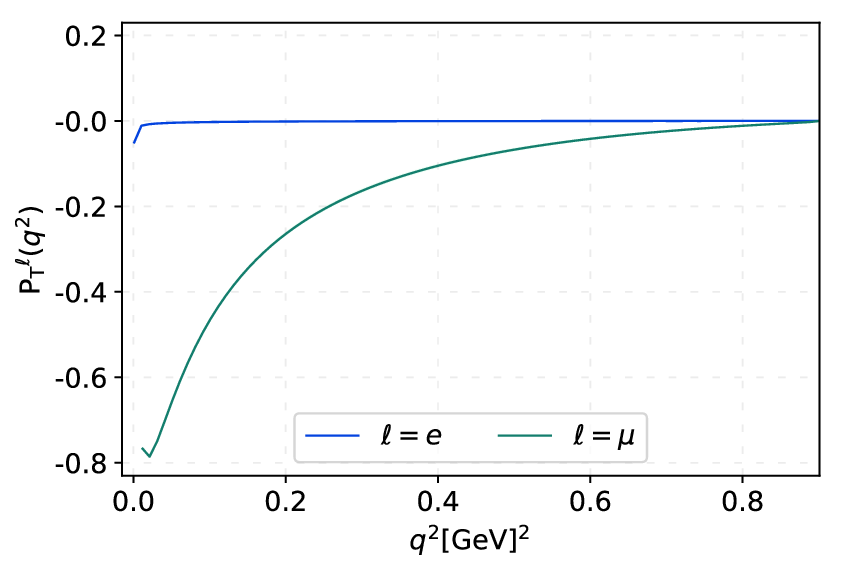}
\end{center}
\caption{The polarization and asymmetry parameters of the decay $D_s^+ \to \phi\ell^+\nu_\ell$, including their central values and uncertainties. The solid lines represent the central values, while the shaded regions indicate the corresponding uncertainties.}
\label{Fig:PSparameters}
\end{figure*}
On this basis, we chose to integrate $q^2$ over the entire physical region from $m_\ell^2$ to $(M_{D_s^+}-m_\phi)^2$, resulting in the following:
\begin{align}
&\Gamma_\mathrm{L} =(16.075_{-2.131}^{+2.494}) \times 10^{-15}~\mathrm{GeV}, \nonumber\\
&\Gamma_\mathrm{T} =(14.837_{-1.240}^{+1.499}) \times 10^{-15}~\mathrm{GeV}, \nonumber\\
&\Gamma_\mathrm{total} =(30.912_{-3.295}^{+3.944}) \times 10^{-15}~\mathrm{GeV}.
\label{eq:widths}
\end{align}

\begin{table}[t]
\footnotesize
\begin{center}
\caption{Comparison of experimental and theoretical predictions on the branching fractions $\mathcal{B}(D_s^+\to\phi \ell^+\nu_\ell)$ with $\ell=e,\mu$ and their corresponding errors (in unit: $10^{-2}$).}
\label{Tab:dB}
\begin{tabular}{l l l}
\hline
\hline
~~~~~~~~~~~~~~~~~~~~~~~~~~~~~~~~~~~~~~&$\mathcal{B}(D_s^+\to\phi e^+\nu_e)$
~~~~~~~~~~~~~~~~~&$\mathcal{B}(D_s^+\to\phi\mu^+\nu_\mu)$        \\
\hline
This Work
&$2.271_{-0.243}^{+0.291}$
&$2.250_{-0.240}^{+0.287}$
\\
HQEFT'06~\cite{Wu:2006rd}
&$2.53_{-0.40}^{+0.37}$
&$2.40_{-0.40}^{+0.35}$
\\
CQM'00~\cite{Melikhov:2000yu}
&$2.57$
&$2.57$
\\
3PSR'03~\cite{Du:2003ja}
&$1.80\pm0.50$
&$-$
\\
CCQM'19~\cite{Ivanov:2019nqd}
&$3.01$
&$2.85$
\\
BaBar'08~\cite{BaBar:2008gpr}
&$1.61\pm0.03\pm0.17$
&$-$
\\
BESIII'17~\cite{BESIII:2017ikf}
&$2.26\pm0.45\pm0.09$
&$1.94\pm0.54$
\\
BESIII'23~\cite{BESIII:2023opt}
&$-$
&$2.25\pm0.09\pm0.07$
\\
PDG~\cite{ParticleDataGroup:2024cfk}
&$2.39\pm0.16$
&$2.24\pm0.11$
\\
\hline
\hline
\end{tabular}
\end{center}
\end{table}
Furthermore, by combining the lifetime of the initial state $D_s^+$-meson $\tau _{D_s^+} = 0.504 ~\mathrm{ps}$ and the CKM matrix element $|V_{cs}| = 0.975$~\cite{ParticleDataGroup:2024cfk}, we can obtain the branching fractions for $D_s^+ \to \phi\ell^+\nu_\ell$ with $\ell = (e,\mu)$, which are shown in Table~\ref{Tab:dB}.
The analysis also incorporates systematic comparisons between experimental measurements and theoretical predictions, {\it i.e.} HQEFT~\cite{Wu:2006rd}, CQM~\cite{Melikhov:2000yu}, 3PSR~\cite{Du:2003ja}, CCQM~\cite{Ivanov:2019nqd}, BaBar~\cite{BaBar:2008gpr}, BESIII~\cite{BESIII:2017ikf,BESIII:2023opt} and PDG~\cite{ParticleDataGroup:2024cfk}.
Significantly, our measured branching fraction for the highly anticipated $\mu$-channel decay shows excellent agreement with BESIII'23~\cite{BESIII:2023opt}, while simultaneously demonstrating good consistency with BESIII'17~\cite{BESIII:2017ikf} in the $e$-channel measurements.
The branching fraction serves as a crucial observable in the SM, where its precise theoretical predictions provide essential parameters for connecting theoretical calculations with experimental measurements.

In the SM, the CKM matrix elements are fundamental parameters whose precise determination is both crucial and indispensable.
Moreover, the structure of the SM requires the CKM matrix element to be unitary.
Therefore, independent precision determinations of CKM matrix elements compared with experimental measurements can test their orthogonality and probe the validity of the SM.
In Table~\ref{Tab:Vcs}, we present the predicted CKM matrix element $|V_{cs}|$ using the branching fraction results $\mathcal{B}(D_s^+\to\phi\mu^+\nu_\mu)=2.25\pm0.09\pm0.07$ from BESIII'23~\cite{BESIII:2023opt}.
Additionally, we have compared our results with those from different research groups, including PDG~\cite{ParticleDataGroup:2024cfk}, BESIII~\cite{BESIII:2018ccy,BESIII:2015tql,BESIII:2015jmz,BESIII:2021anh},
HFLAV~\cite{HeavyFlavorAveragingGroupHFLAV:2024ctg}, Bayesian~\cite{Bolognani:2024cmr}, HPQCD~\cite{Chakraborty:2021qav}, LQCD~\cite{Riggio:2017zwh}, Belle~\cite{Belle:2013isi} and CKMfitter~\cite{Charles:2004jd}.
As shown, our results demonstrate excellent agreement with the PDG averages~\cite{ParticleDataGroup:2024cfk}, while also showing remarkable consistency with the BESIII(I)'15~\cite{BESIII:2015tql} experimental measurements.
In comparisons with theoretical predictions, our results show close agreement with LQCD'17~\cite{Riggio:2017zwh}  calculations within uncertainty ranges.
\begin{table}[t]
\footnotesize
\begin{center}
\caption{The results of $|V_{cs}|$ in different channels are compared, including the results of the experimental groups and the theoretical group. }
\label{Tab:Vcs}
\begin{tabular}{l l}
\hline
\hline
~~~~~~~~~~~~~~~~~~~~~~~~~~~~~~~~~~~~~~~~~~~~~~~~~~~~~~~&$|V_{cs}|$
\\
\hline
This Work                &$0.975_{-0.066}^{+0.067}$
\\
PDG~\cite{ParticleDataGroup:2024cfk}   &$0.975\pm0.006$
\\
HFLAV'24~\cite{HeavyFlavorAveragingGroupHFLAV:2024ctg}  &$0.9808\pm0.0079\pm0.0020$
\\
Bayesian'24~\cite{Bolognani:2024cmr}    &$0.957\pm0.003$
\\
HPQCD'21~\cite{Chakraborty:2021qav}     &$0.9663(80)$
\\
LQCD'17~\cite{Riggio:2017zwh}           &$0.970(33)$
\\
Belle'13~\cite{Belle:2013isi}        &$0.97341(22)$
\\
CKMfitter'05~\cite{Charles:2004jd}  &$1.04\pm0.16$
\\
BESIII(I)'15~\cite{BESIII:2015tql}          &$0.9601\pm0.0033\pm0.0047\pm0.0239$
\\
BESIII(II)'15~\cite{BESIII:2015jmz}      &$0.975\pm0.008\pm0.015\pm0.025$
\\
BESIII'18~\cite{BESIII:2018ccy}         &$0.955\pm0.005\pm0.004\pm0.024$
\\
BEIII'21~\cite{BESIII:2021anh}       &$0.973\pm0.009\pm0.014$
\\
\hline
\hline
\end{tabular}
\end{center}
\end{table}
Discrepancies remain between our predictions and other theoretical groups due to methodological differences, originating from distinct technical approaches. This underscores the need for further investigation of $s\bar{s}$-composition mesons to achieve higher precision in $|V_{cs}|$ determinations.

Meanwhile, one presents predictions for the polarization parameters and asymmetry parameters $A_\mathrm{FB}^\ell$, $C_\mathrm{F}^\ell$, $P_\mathrm{L}^\ell$, $P_\mathrm{T}^\ell$ and $F_\mathrm{L}^\ell$ of the $D_s^+\to\phi$ decay, providing more detailed information for hadronic semileptonic decays in Fig.~\ref{Fig:PSparameters}, whose errors caused by different choices of input parameters are shown by shaded bands. In Fig.~\ref{Fig:PSparameters}, one observe distinct behaviors in the polarization and asymmetry parameters between electrons and muons in the low $q^2$-region. This primarily arises because the electrons negligible mass renders its polarization effects less pronounced in interactions, whereas the muons larger mass introduces more complex interaction mechanisms, leading to unique polarization and asymmetry patterns. Additionally, the Fig.~\ref{Fig:PSparameters} reveal that all uncertainties are remarkably small, particularly for observables like the longitudinal or transverse polarization fractions $(P_\mathrm{L}^\ell,P_\mathrm{T}^\ell)$ of the final-state charged lepton, where the uncertainties are practically negligible.

After performing the integration, one obtains the integrated value of these observables, which are presented in Table~\ref{Tab:Psvalue} and compared with the results from the CCQM~\cite{Ivanov:2019nqd} and the RQM~\cite{Faustov:2019mqr}. The polarization and asymmetry parameters exhibit distinct variations depending on the lepton mass. Our predicted values are systematically slightly lower than those of the RQM and CCQM but remain entirely within the ranges predicted by these models.
\begin{table}[t]
\footnotesize
\begin{center}
\caption{The polarization and asymmetry parameters of the decay $D_s^+\to\phi\ell^+\nu_\ell$ are calculated with all input parameters set to their central values.}
\label{Tab:Psvalue}
\begin{tabular}{p{2cm} p{1.2cm} p{1.2cm} p{1.2cm} p{1.2cm} p{0.8cm}}
\hline
\hline
 &$A_\mathrm{FB}^e$ &$A_\mathrm{FB}^\mu$
&$C_\mathrm{F}^e$ &$C_\mathrm{F}^\mu$
&$P_\mathrm{L}^e$
\\
\hline
This work     &$-0.174$ &$-0.203$ &$-0.393$ &$-0.295$&$0.900$
\\
CCQM'19~\cite{Ivanov:2019nqd}    &$-0.18$ &$-0.21$ &$-0.43$ &$-0.34$&$1.00$
\\
RQM'20~\cite{Faustov:2019mqr}    &$-0.21$ &$-0.24$ &$-0.49$ &$-0.35$&$1.00$
\\
\hline
&$P_\mathrm{L}^\mu$
&$P_\mathrm{T}^e$ &$P_\mathrm{T}^\mu$
&$F_\mathrm{L}^e$ &$F_\mathrm{L}^\mu$
\\
This work &$0.781$ &$-0.001$ &$-0.145$ &$0.475$ &$0.471$
\\
CCQM'19~\cite{Ivanov:2019nqd} &$0.91$ &$-0.11$ &$-0.14$ &$0.53$ &$0.50$
\\
RQM'20~\cite{Faustov:2019mqr}  &$0.90$ &$0.00$ &$-0.15$ &$0.54$ &$0.54$
\\
\hline
\hline
\end{tabular}
\end{center}
\end{table}

\section{Summary}\label{sec:summary}
In this work, we employed the LCHO model to construct the transverse LCDA of the $\phi$-meson, and performed LCSR calculations for the semileptonic decay $D_s^+\to\phi\ell^+\nu_\ell$ using right-handed chiral correlator. In our calculations, we proposed an improved transverse twist-2 LCDA $\phi_{2;\phi}^\bot(x,\mu_0)$ for the $\phi$-meson within the LCHO model framework, with its behavior presented in Fig.~\ref{Fig:DA}. Our model exhibits a double-peak behavior, showing strong similarity to the conventional SR predictions. In our study of $D_s^+\to\phi$ decays, we employed right-handed chiral correlator in the LCSR framework to isolate distinct twist contributions. The TFFs are calculated at the large recoil point $q^2=0$ and have been given in Table~\ref{Tab:TFFs}. Through those TFFs, we systematically calculated multiple physical observables including the branching fractions, the decay widths, and the polarization and asymmetry parameters. We have presented the differential decay width of the semileptonic decay $D_s^+\to \phi\ell^+\nu_\ell$ with $\ell=(e,\mu)$ in Fig.~\ref{Fig:dGamma} and the branching fractions in Table~\ref{Tab:dB}. For both $e$ and $\mu$ channels, our theoretical predictions show full consistency with experimental measurements within quoted uncertainties. Additionally, we extracted the CKM matrix element $|V_{cs}|$ from our analysis, with the determined value presented in Table~\ref{Tab:Vcs}. Our determined $|V_{cs}|$ value shows agreement with experimental measurements and is consistent with the PDG average within uncertainties. Furthermore, we have calculated the longitudinal, transverse, and total decay widths, with the results presented in Eq.~\eqref{eq:widths}. We further predicted the forward-backward asymmetries, the lepton-side convexity parameters, as well as lepton and vector meson longitudinal and polarization parameters, which have been showed in Table~\ref{Tab:Psvalue} and Fig.~\ref{Fig:PSparameters}. These observables are experimentally measurable in the near future, and their measurements will provide crucial tests for the validity of our $\phi$-meson LCDA model.

\acknowledgments
Tao Zhong and Hai-Bing Fu would like to thank the Institute of High Energy Physics of Chinese Academy of Sciences for their warm and kind hospitality. This work was supported in part by the National Natural Science Foundation of China under Grant No.12265009, No.12265010, the Project of Guizhou Provincial Department of Science and Technology under Grant No.MS[2025]219, No.CXTD[2025]030, No.ZK[2023]024.


\begin{thebibliography}{99}

\bibitem{Kobayashi:1973fv}
M.~Kobayashi and T.~Maskawa,
CP Violation in the Renormalizable Theory of Weak Interaction,
\href{https://academic.oup.com/ptp/article/49/2/652/1858101?login=true}
{Prog. Theor. Phys. \textbf{49}, 652-657 (1973)}.

\bibitem{Cabibbo:1963yz}
N.~Cabibbo,
Unitary Symmetry and Leptonic Decays,
\href{https://journals.aps.org/prl/abstract/10.1103/PhysRevLett.10.531}
{Phys. Rev. Lett. \textbf{10}, 531-533 (1963)}.

\bibitem{Wolfenstein:1983yz}
L.~Wolfenstein,
Parametrization of the Kobayashi-Maskawa Matrix,
\href{https://journals.aps.org/prl/abstract/10.1103/PhysRevLett.51.1945}
{Phys. Rev. Lett. \textbf{51}, 1945 (1983)}.

\bibitem{BESIII:2018nzb}
M.~Ablikim \textit{et al.} [BESIII Collaboration],
Measurement of the branching fraction for the semi-leptonic decay $D^{0(+)}\to \pi^{-(0)}\mu^+\nu_\mu$ and test of lepton universality,
\href{https://journals.aps.org/prl/abstract/10.1103/PhysRevLett.121.171803}
{Phys. Rev. Lett. \textbf{121}, no.17, 171803 (2018)}.
[\href{https://arxiv.org/abs/1802.05492}
{arXiv:1802.05492}]

\bibitem{BESIII:2018ccy}
M.~Ablikim \textit{et al.} [BESIII Collaboration],
Study of the $D^0\to K^-\mu^+\nu_\mu$ dynamics and test of lepton flavor universality with $D^0\to K^-\ell^+\nu_\ell$ decays,
\href{https://journals.aps.org/prl/abstract/10.1103/PhysRevLett.122.011804}
{Phys. Rev. Lett. \textbf{122}, no.1, 011804 (2019)}.
[\href{https://arxiv.org/abs/1810.03127}
{arXiv:1810.03127}]

\bibitem{Ablikim:2020hsc}
M.~Ablikim [BESIII Collaboration],
First Observation of $D^+ \rightarrow \eta\mu^+\nu_\mu$ and Measurement of Its Decay Dynamics,
\href{https://journals.aps.org/prl/abstract/10.1103/PhysRevLett.124.231801}
{Phys. Rev. Lett. \textbf{124}, no.23, 231801 (2020)}.
[\href{https://arxiv.org/abs/2003.12220}
{arXiv:2003.12220}]

\bibitem{BESIII:2019qci}
M.~Ablikim \textit{et al.} [BESIII Collaboration],
Measurement of the Dynamics of the Decays $D_s^+ \rightarrow \eta^{(\prime)} e^+ \nu_e$,
\href{https://journals.aps.org/prl/abstract/10.1103/PhysRevLett.122.121801}
{Phys. Rev. Lett. \textbf{122}, no.12, 121801 (2019)}.
[\href{https://arxiv.org/abs/1901.02133}
{arXiv:1901.02133}]

\bibitem{BESIII:2017ikf}
M.~Ablikim \textit{et al.} [BESIII Collaboration],
Measurements of the branching fractions for the semi-leptonic decays $D^+_s\to\phi e^{+}\nu_{e}$, $\phi \mu^{+}\nu_\mu $, $\eta \mu^{+}\nu_\mu $ and $\eta'\mu^{+}\nu_\mu $,
\href{https://journals.aps.org/prd/abstract/10.1103/PhysRevD.97.012006}
{Phys. Rev. D \textbf{97}, no.1, 012006 (2018)}.
[\href{https://arxiv.org/abs/1709.03680}
{arXiv:1709.03680}]

\bibitem{Ke:2023qzc}
B.~C.~Ke, J.~Koponen, H.~B.~Li and Y.~Zheng,
\href{https://www.annualreviews.org/content/journals/10.1146/annurev-nucl-110222-044046}
{Ann. Rev. Nucl. Part. Sci. \textbf{73}, 285-314 (2023)}.
[\href{https://arxiv.org/abs/2310.05228}
{arXiv:2310.05228}]

\bibitem{BESIII:2021ynj}
M.~Ablikim \textit{et al.} [BESIII Collaboration],
First Measurement of the Absolute Branching Fraction of $\Lambda \to p \mu^- \bar{\nu}_\mu $,
\href{https://journals.aps.org/prl/abstract/10.1103/PhysRevLett.127.121802}
{Phys. Rev. Lett. \textbf{127}, no.12, 121802 (2021)}.
[\href{https://arxiv.org/abs/2107.06704 }
{arXiv:2107.06704}]

\bibitem{Belle:2021crz}
Y.~B.~Li \textit{et al.} [Belle Collaboration],
Measurements of the branching fractions of the semileptonic decays $\Xi_{c}^{0} \to \Xi^{-} \ell^{+} \nu_{\ell}$ and the asymmetry parameter of $\Xi_{c}^{0} \to \Xi^{-} \pi^{+}$,
\href{https://journals.aps.org/prl/abstract/10.1103/PhysRevLett.127.121803}
{Phys. Rev. Lett. \textbf{127}, no.12, 121803 (2021)}.
[\href{https://arxiv.org/abs/2103.06496}
{arXiv:2103.06496}]

\bibitem{Belle:2021dgc}
Y.~B.~Li \textit{et al.} [Belle Collaboration],
First test of lepton flavor universality in the charmed baryon decays $\Omega_c^0\to\Omega^-\ell^+\nu_\ell$ using data of the Belle experiment,
\href{https://journals.aps.org/prd/abstract/10.1103/PhysRevD.105.L091101}
{Phys. Rev. D \textbf{105}, no.9, L091101 (2022)}.
[\href{https://arxiv.org/abs/2112.10367}
{arXiv:2112.10367}]

\bibitem{BESIII:2015ysy}
M.~Ablikim \textit{et al.} [BESIII],
Measurement of the absolute branching fraction for $\Lambda^+_{c}\to \Lambda e^+\nu_e$,
\href{https://journals.aps.org/prl/abstract/10.1103/PhysRevLett.115.221805}
{Phys. Rev. Lett. \textbf{115}, no.22, 221805 (2015)}.
[\href{https://arxiv.org/abs/1510.02610}
{arXiv:1510.02610}]

\bibitem{Li:2020ylu}
H.~B.~Li and M.~Z.~Yang,
Semileptonic decay of $D_s^+\to \pi^0 \ell^+ \nu_\ell$ via neutral meson mixing,
\href{https://www.sciencedirect.com/science/article/pii/S0370269320306821?via\%3Dihub}
{Phys. Lett. B \textbf{811}, 135879 (2020)}.
[\href{https://arxiv.org/abs/2006.15798}
{arXiv:2006.15798}]

\bibitem{Benayoun:2001qz}
M.~Benayoun and H.~B.~O'Connell,
Isospin symmetry breaking within the HLS model: A Full ($\rho$, $\omega$, $\phi$) mixing scheme,
\href{https://link.springer.com/article/10.1007/s100520100806}
{Eur. Phys. J. C \textbf{22}, 503-520 (2001)}.
[\href{https://arxiv.org/abs/nucl-th/0107047}
{arXiv:nucl-th/0107047}]

\bibitem{Benayoun:2007cu}
M.~Benayoun, P.~David, L.~DelBuono, O.~Leitner and H.~B.~O'Connell,
The Dipion Mass Spectrum In $e^+ e^-$ Annihilation and tau Decay: A Dynamical ($\rho, \omega, \phi$) Mixing Approach,
\href{https://link.springer.com/article/10.1140/epjc/s10052-008-0586-6}
{Eur. Phys. J. C \textbf{55}, 199-236 (2008)}.
[\href{https://arxiv.org/abs/0711.4482}
{arXiv:0711.4482}]

\bibitem{Benayoun:2008tm}
M.~Benayoun,
Effects of the ($\rho, \omega, \phi$) mixing on the dipion mass spectrum in $e^+ e^-$ annihilation and $\tau$ decay,
[\href{https://arxiv.org/abs/0805.1835}
{arXiv:0805.1835}].



\bibitem{Qian:2009dc}
W.~Qian and B.~Q.~Ma,
Tri-meson-mixing of $\pi$-$\eta$-$\eta'$ and $\rho$-$\omega$-$\phi$ in the light-cone quark model,''
\href{https://link.springer.com/article/10.1140/epjc/s10052-009-1220-y}
{Eur. Phys. J. C \textbf{65}, 457-465 (2010)}.
[\href{https://arxiv.org/abs/0912.0612}
{arXiv:0912.0612}]

\bibitem{Fritzsch:2001aj}
H.~Fritzsch,
The Breaking of isospin and the rho omega system,
[\href{https://arxiv.org/abs/hep-ph/0106273}
{arXiv:hep-ph/0106273}].


\bibitem{Shakin:1995md}
C.~M.~Shakin and W.~D.~Sun,
Microscopic foundations of the vector meson dominance model via the momentum space bosonization of an extended NJL model,
\href{https://journals.aps.org/prd/abstract/10.1103/PhysRevD.55.2874}
{Phys. Rev. D \textbf{55}, 2874-2888 (1997)}.

\bibitem{Mitchell:1996dn}
K.~L.~Mitchell and P.~C.~Tandy,
Pion loop contribution to rho - omega mixing and mass splitting,
\href{https://journals.aps.org/prc/abstract/10.1103/PhysRevC.55.1477}
{Phys. Rev. C \textbf{55}, 1477-1491 (1997)}.
[\href{https://arxiv.org/abs/nucl-th/9607025}
{arXiv:nucl-th/9607025}]

\bibitem{OConnell:1997ggd}
H.~B.~O'Connell, A.~W.~Thomas and A.~G.~Williams,
Extracting the $\rho - \omega$ mixing amplitude from the pion form-factor,
\href{https://www.sciencedirect.com/science/article/abs/pii/S0375947497884254?via\%3Dihub}
{Nucl. Phys. A \textbf{623}, 559-569 (1997)}.
[\href{https://arxiv.org/abs/hep-ph/9703248}
{arXiv:hep-ph/9703248}]

\bibitem{Wang:2000gq}
X.~J.~Wang and M.~L.~Yan,
Isospin breaking and omega ---{\ensuremath{>}} pi+ pi- decay,
\href{https://journals.aps.org/prd/abstract/10.1103/PhysRevD.62.094013}
{Phys. Rev. D \textbf{62}, 094013 (2000)}.
[\href{https://arxiv.org/abs/hep-ph/0003218}
{arXiv:hep-ph/0003218}]

\bibitem{Gao:1998gr}
D.~N.~Gao and M.~L.~Yan,
Rho0 - omega mixing in U(3)-L x U(3)-R chiral theory of mesons,
\href{https://link.springer.com/article/10.1007/s100500050180}
{Eur. Phys. J. A \textbf{3}, 293-298 (1998)}.
[\href{https://arxiv.org/abs/hep-ph/9801210}
{arXiv:hep-ph/9801210}]

\bibitem{Fritzsch:2000pg}
H.~Fritzsch and A.~S.~Muller,
Isospin symmetry breaking and the rho omega system,
\href{https://www.sciencedirect.com/science/article/abs/pii/S0920563201011410?via\%3Dihub}
{Nucl. Phys. B Proc. Suppl. \textbf{96}, 273-276 (2001)}.
[\href{https://arxiv.org/abs/hep-ph/0011278}
{arXiv:hep-ph/0011278}]




\bibitem{Li:2019xwh}
S.~T.~Li and G.~L{\"u},
Direct $CP$ violation for $\bar{B}_s^0 \to \phi {\pi^+}{\pi^-}$ in Perturbative QCD,
\href{https://journals.aps.org/prd/abstract/10.1103/PhysRevD.99.116009}
{Phys. Rev. D \textbf{99}, no.11, 116009 (2019)}.
[\href{https://arxiv.org/abs/1904.11824}
{arXiv:1904.11824}]

\bibitem{OConnell:1995nse}
H.~B.~O'Connell, B.~C.~Pearce, A.~W.~Thomas and A.~G.~Williams,
$\rho - \omega$ mixing, vector meson dominance and the pion form-factor,
\href{https://www.sciencedirect.com/science/article/abs/pii/S0146641097000446?via\%3Dihub}
{Prog. Part. Nucl. Phys. \textbf{39}, 201-252 (1997)}.
[\href{https://arxiv.org/abs/hep-ph/9501251}
{arXiv:hep-ph/9501251}]


\bibitem{OConnell:1994czf}
H.~B.~O'Connell, B.~C.~Pearce, A.~W.~Thomas and A.~G.~Williams,
Constraints on the momentum dependence of $\rho - \omega$ mixing,
\href{https://www.sciencedirect.com/science/article/abs/pii/0370269394009910?via\%3Dihub}
{Phys. Lett. B \textbf{336}, 1-5 (1994)}.
[\href{https://arxiv.org/abs/hep-ph/9405273}
{arXiv:hep-ph/9405273}]


\bibitem{OConnell:1996amv}
H.~B.~O'Connell,
Recent developments in $\rho - \omega$ mixing,
\href{https://www.publish.csiro.au/ph/P96041}
{Austral. J. Phys. \textbf{50}, 255-262 (1997)}.
[\href{https://arxiv.org/abs/hep-ph/9604375}
{arXiv:hep-ph/9604375}]


\bibitem{Benayoun:1999fv}
M.~Benayoun, L.~DelBuono, S.~Eidelman, V.~N.~Ivanchenko and H.~B.~O'Connell,
Radiative decays, nonet symmetry and SU(3) breaking,
\href{https://link.aps.org/doi/10.1103/PhysRevD.59.114027}
{Phys. Rev. D \textbf{59}, 114027 (1999)}.
[\href{https://arxiv.org/abs/hep-ph/9902326}
{hep-ph/9902326}]

\bibitem{Gronau:2008kk}
M.~Gronau and J.~L.~Rosner,
$B$ decays dominated by $\omega^- \phi$ mixing,
\href{https://www.sciencedirect.com/science/article/pii/S0370269308008423?via\%3Dihub}
{Phys. Lett. B \textbf{666}, 185-188 (2008)}.
[\href{https://arxiv.org/abs/0806.3584}
{arXiv:0806.3584}]

\bibitem{Gronau:2009mp}
M.~Gronau and J.~L.~Rosner,
$\omega - \phi$ mixing and weak annihilation in $D_s$ decays,
\href{https://link.aps.org/doi/10.1103/PhysRevD.79.074006}
{Phys. Rev. D \textbf{79}, 074006 (2009)}.
[\href{https://arxiv.org/abs/0902.1363}
{arXiv:0902.1363}]

\bibitem{Kucukarslan:2006wk}
A.~Kucukarslan and U.~G.~Meissner,
$\omega-\phi$ mixing in chiral perturbation theory,
\href{https://www.worldscientific.com/doi/abs/10.1142/S0217732306020743}
{Mod. Phys. Lett. A \textbf{21}, 1423-1430 (2006)}.
[\href{https://arxiv.org/abs/hep-ph/0603061}
{arXiv:hep-ph/0603061}]

\bibitem{Yang:2025qgh}
Y.~L.~Yang, Y.~L.~Song, F.~P.~Peng, H.~B.~Fu, T.~Zhong and S.~Ullah,
Exploring exclusive decay $B^+\to \omega\ell^+\nu$ within LCSR,
[\href{https://arxiv.org/abs/2504.05650}
{arXiv:2504.05650}].

\bibitem{Ambrosino:2009sc}
F.~Ambrosino, A.~Antonelli, M.~Antonelli, F.~Archilli, P.~Beltrame, G.~Bencivenni, S.~Bertolucci, C.~Bini, C.~Bloise and S.~Bocchetta, \textit{et al.}
A Global fit to determine the pseudoscalar mixing angle and the gluonium content of the $\eta^\prime$ meson,
\href{https://iopscience.iop.org/article/10.1088/1126-6708/2009/07/105}
{JHEP \textbf{07}, 105 (2009)}.
[\href{https://arxiv.org/abs/0906.3819}
{arXiv:0906.3819}]

\bibitem{Huang:2021kfm}
Q.~Huang, J.~Z.~Wang, R.~G.~Ping and X.~Liu,
Detecting the polarization in $\chi_{cJ} \to \phi \phi $ decays to probe hadronic loop effect,
\href{https://journals.aps.org/prd/abstract/10.1103/PhysRevD.103.096006}
{Phys. Rev. D \textbf{103}, no.9, 096006 (2021)}.
[\href{https://arxiv.org/abs/2102.07104}
{arXiv:2102.07104}]

\bibitem{Volkov:2020jor}
M.~K.~Volkov, A.~A.~Pivovarov and K.~Nurlan,
On the mixing angle of the vector mesons $\omega(782)$ and $\phi(1020)$,
\href{https://www.worldscientific.com/doi/abs/10.1142/S0217732320502004}
{Mod. Phys. Lett. A \textbf{35}, no.24, 2050200 (2020)}.
[\href{https://arxiv.org/abs/2005.00763}
{arXiv:2005.00763}]

\bibitem{Bramon:1994cb}
A.~Bramon, A.~Grau and G.~Pancheri,
Effective chiral lagrangians with an SU(3) broken vector meson sector,
\href{https://www.sciencedirect.com/science/article/abs/pii/037026939401625M?via\%3Dihub}
{Phys. Lett. B \textbf{345}, 263-268 (1995)}.
[\href{https://arxiv.org/abs/hep-ph/9411269}
{arXiv:hep-ph/9411269}]

\bibitem{Bramon:1994pq}
A.~Bramon, A.~Grau and G.~Pancheri,
Radiative vector meson decays in SU(3) broken effective chiral Lagrangians,
\href{https://www.sciencedirect.com/science/article/abs/pii/037026939401543L?via\%3Dihub}
{Phys. Lett. B \textbf{344}, 240-244 (1995)}.

\bibitem{Benayoun:2000ti}
M.~Benayoun, L.~DelBuono, P.~Leruste and H.~B.~O'Connell,
An Effective approach to VMD at one loop order and the departures from ideal mixing for vector mesons,
\href{https://link.springer.com/article/10.1007/s100520000463}
{Eur. Phys. J. C \textbf{17}, 303-321 (2000)}.
[\href{https://arxiv.org/abs/nucl-th/0004005}
{arXiv:nucl-th/0004005}]

\bibitem{BESIII:2023opt}
M.~Ablikim \textit{et al.} [BESIII Collaboration],
Studies of the decay $ {\textrm{D}}_{\textrm{s}}^{+}\to {\textrm{K}}^{+}{\textrm{K}}^{-}{\mu}^{+}{\nu}_{\mu } $,
\href{https://link.springer.com/article/10.1007/JHEP12(2023)072}
{JHEP \textbf{12}, 072 (2023)}.
[\href{https://arxiv.org/abs/2307.03024}
{arXiv:2307.03024}]


\bibitem{BaBar:2008gpr}
B.~Aubert \textit{et al.} [BaBar Collaboration],
Study of the decay $D^+_{s} \to K^{+} K^{-} e^{+} \nu_{e}$,
\href{https://journals.aps.org/prd/abstract/10.1103/PhysRevD.78.051101}
{Phys. Rev. D \textbf{78}, 051101 (2008)}.
[\href{https://arxiv.org/abs/0807.1599}
{arXiv:0807.1599}]

\bibitem{Hietala:2015jqa}
J.~Hietala, D.~Cronin-Hennessy, T.~Pedlar and I.~Shipsey,
Exclusive $D_s$ semileptonic branching fraction measurements,
\href{https://journals.aps.org/prd/abstract/10.1103/PhysRevD.92.012009}
{Phys. Rev. D \textbf{92}, no.1, 012009 (2015)}.
[\href{https://arxiv.org/abs/1505.04205}
{arXiv:1505.04205}]

\bibitem{Wu:2006rd}
Y.~L.~Wu, M.~Zhong and Y.~B.~Zuo,
$B_{(s)}$, $D_{(s)}\to \pi$, $K$, $\eta$, $\rho$, $K^*$, $\omega$, $\phi$ Transition Form Factors and Decay Rates with Extraction of the CKM parameters $|V_{ub}|$, $|V_{cs}|$, $|V_{cd}|$,
\href{https://www.worldscientific.com/doi/abs/10.1142/S0217751X06033209}
{Int. J. Mod. Phys. A \textbf{21}, 6125-6172 (2006)}
[\href{https://arxiv.org/abs/hep-ph/0604007}
{hep-ph/0604007}]



\bibitem{Fajfer:2005ug}
S.~Fajfer and J.~F.~Kamenik,
Charm meson resonances and $D \to V$ semileptonic form-factors,
\href{https://journals.aps.org/prd/abstract/10.1103/PhysRevD.72.034029}
{Phys. Rev. D \textbf{72}, 034029 (2005)}.
[\href{https://arxiv.org/abs/hep-ph/0506051}
{hep-ph/0506051}]

\bibitem{Melikhov:2000yu}
D.~Melikhov and B.~Stech,
Weak form-factors for heavy meson decays: An Update,
\href{https://journals.aps.org/prd/abstract/10.1103/PhysRevD.62.014006}
{Phys. Rev. D \textbf{62}, 014006 (2000)}.
[\href{https://arxiv.org/abs/hep-ph/0001113}
{hep-ph/0001113}]

\bibitem{Soni:2019huk}
N.~R.~Soni and J.~N.~Pandya,
Decay $D^+_s \to \phi\ell^+\nu_{\ell}$ in Covariant Quark Model,
\href{https://www.epj-conferences.org/articles/epjconf
/abs/2019/07/epjconf_charm2018_06010/epjconf_charm2018_06010.html}
{EPJ Web Conf. \textbf{202}, 06010 (2019)}.

\bibitem{Du:2003ja}
D.~S.~Du, J.~W.~Li and M.~Z.~Yang,
Form-factors and semileptonic decay of $D^+_s \to\phi\bar{l}\nu$ from QCD sum rule,
\href{https://link.springer.com/article/10.1140/epjc/s2004-01979-9}
{Eur. Phys. J. C \textbf{37}, no.2, 173-184 (2004)}.
[\href{https://arxiv.org/abs/hep-ph/0308259}
{hep-ph/0308259}]

\bibitem{Bediaga:2003hr}
I.~Bediaga and M.~Nielsen,
$D_{(s)}$ decays into $\phi$ and $f_{0}(980)$ mesons,
\href{https://journals.aps.org/prd/abstract/10.1103/PhysRevD.68.036001}
{Phys. Rev. D \textbf{68}, 036001 (2003)}.
[\href{https://arxiv.org/abs/hep-ph/0304193}
{hep-ph/0304193}].

\bibitem{Verma:2011yw}
R.~C.~Verma,
Decay constants and form factors of s-wave and p-wave mesons in the covariant light-front quark model,
\href{https://iopscience.iop.org/article/10.1088/0954-3899/39/2/025005}
{J. Phys. G \textbf{39}, 025005 (2012)}.
[\href{https://arxiv.org/abs/1103.2973}
{arXiv:1103.2973}]

\bibitem{Cheng:2017pcq}
H.~Y.~Cheng and X.~W.~Kang,
Branching fractions of semileptonic $D$ and $D_s$ decays from the covariant light-front qluark model,
\href{https://link.springer.com/article/10.1140/epjc/s10052-017-5170-5}
{Eur. Phys. J. C \textbf{77}, no.9, 587 (2017)}.
[\href{https://arxiv.org/abs/1707.02851}
{arXiv:1707.02851}]

\bibitem{Chang:2019mmh}
Q.~Chang, X.~N.~Li and L.~T.~Wang,
Revisiting the form factors of $P\rightarrow V$ transition within the light-front quark models,
\href{https://link.springer.com/article/10.1140/epjc/s10052-019-6949-3}
{Eur. Phys. J. C \textbf{79}, no.5, 422 (2019)}.
[\href{https://arxiv.org/abs/1905.05098}
{arXiv:1905.05098}]

\bibitem{Aliev:2004vf}
T.~M.~Aliev, M.~Savci and A.~Ozpineci,
Form factors of $D_s^+ \to \phi \bar{\ell} \nu$ decay in QCD light cone sum rule,
\href{https://link.springer.com/article/10.1140/epjc/s2004-02034-9}
{Eur. Phys. J. C \textbf{38}, 85-91 (2004)}.

\bibitem{Ivanov:2019nqd}
M.~A.~Ivanov, J.~G.~K\"orner, J.~N.~Pandya, P.~Santorelli, N.~R.~Soni and C.~T.~Tran,
Exclusive semileptonic decays of D and D$_{s}$ mesons in the covariant confining quark model,
\href{https://link.springer.com/article/10.1007/s11467-019-0908-1}
{Front. Phys. (Beijing) \textbf{14}, no.6, 64401 (2019)}.
[\href{https://arxiv.org/abs/1904.07740}
{arXiv:1904.07740}]

\bibitem{Donald:2013pea}
G.~C.~Donald \textit{et al.} [HPQCD],
$V_{cs}$ from $D_s \to \phi \ell \nu$ semileptonic decay and full lattice QCD,
\href{https://journals.aps.org/prd/abstract/10.1103/PhysRevD.90.074506}
{Phys. Rev. D \textbf{90}, no.7, 074506 (2014)}.
[\href{https://arxiv.org/abs/1311.6669}
{arXiv:1311.6669}]

\bibitem{Donald:2011ff}
G.~Donald, C.~Davies and J.~Koponen,
Axial vector form factors in $D_s \to \phi$ semileptonic decays from lattice QCD,
\href{https://pos.sissa.it/139/278}
{PoS \textbf{LATTICE2011}, 278 (2011)}.
[\href{https://arxiv.org/abs/1111.0254}
{arXiv:1111.0254}]

\bibitem{Koponen:2011ev}
J.~Koponen \textit{et al.} [HPQCD],
The $D$ $\to$ $K$ and $D$ $\to$ $\pi$ semileptonic decay form factors from Lattice QCD,
\href{https://pos.sissa.it/139/286}
{PoS \textbf{LATTICE2011}, 286 (2011)}.
[\href{https://arxiv.org/abs/1111.0225}
{arXiv:1111.0225}]

\bibitem{Faustov:2019mqr}
R.~N.~Faustov, V.~O.~Galkin and X.~W.~Kang,
Semileptonic decays of $D$ and $D_s$ mesons in the relativistic quark model,
\href{https://journals.aps.org/prd/abstract/10.1103/PhysRevD.101.013004}
{Phys. Rev. D \textbf{101}, no.1, 013004 (2020)}.
[\href{https://arxiv.org/abs/1911.08209}
{arXiv:1911.08209}]

\bibitem{Faustov:2013ima}
R.~N.~Faustov and V.~O.~Galkin,
Charmless weak $B_s$ decays in the relativistic quark model,
\href{https://journals.aps.org/prd/abstract/10.1103/PhysRevD.87.094028}
{Phys. Rev. D \textbf{87}, no.9, 094028 (2013)}.
[\href{https://arxiv.org/abs/1304.3255}
{arXiv:1304.3255}]

\bibitem{Faustov:2012mt}
R.~N.~Faustov and V.~O.~Galkin,
Weak decays of $B_s$ mesons to $D_s$ mesons in the relativistic quark model,
\href{https://journals.aps.org/prd/abstract/10.1103/PhysRevD.87.034033}
{Phys. Rev. D \textbf{87}, no.3, 034033 (2013)}.
[\href{https://arxiv.org/abs/1212.3167}
{arXiv:1212.3167}]

\bibitem{Ebert:2009ua}
D.~Ebert, R.~N.~Faustov and V.~O.~Galkin,
Heavy-light meson spectroscopy and Regge trajectories in the relativistic quark model,
\href{https://link.springer.com/article/10.1140/epjc/s10052-010-1233-6}
{Eur. Phys. J. C \textbf{66}, 197-206 (2010)}.
[\href{https://arxiv.org/abs/0910.5612}
{arXiv:0910.5612}]

\bibitem{Hussain:1995jq}
F.~Hussain, A.~N.~Ivanov and N.~I.~Troitskaya,
On the form-factors of the $D_s^+ \to \phi \mu^+\nu_\mu$ decay,
\href{https://www.sciencedirect.com/science/article/abs/pii/0370269395015329?via\%3Dihub}
{Phys. Lett. B \textbf{369}, 351-357 (1996)}.
[\href{https://arxiv.org/abs/hep-ph/9505273}
{hep-ph/9505273}]

\bibitem{Sekihara:2015iha}
T.~Sekihara and E.~Oset,
Investigating the nature of light scalar mesons with semileptonic decays of D mesons,
\href{https://journals.aps.org/prd/abstract/10.1103/PhysRevD.92.054038}
{Phys. Rev. D \textbf{92}, no.5, 054038 (2015)}.
[\href{https://arxiv.org/abs/1507.02026}
{arXiv:1507.02026}]

\bibitem{Fu:2014uea}
H.~B.~Fu, X.~G.~Wu and Y.~Ma,
$B\to K^*$ Transition Form Factors and the Semi-leptonic Decay $B \to K^* \mu^+ \mu^-$,
\href{https://iopscience.iop.org/article/10.1088/0954-3899/43/1/015002}
{J. Phys. G \textbf{43}, no.1, 015002 (2016)}.
[\href{https://arxiv.org/abs/1411.6423}
{arXiv:1411.6423}]

\bibitem{Ball:2004rg}
P.~Ball and R.~Zwicky,
$B_{d,s} \to  \rho, \omega, K^*, \phi$ decay form-factors from light-cone sum rules revisited,
\href{https://journals.aps.org/prd/abstract/10.1103/PhysRevD.71.014029}
{Phys. Rev. D \textbf{71}, 014029 (2005)}.
[\href{https://arxiv.org/abs/hep-ph/0412079}
{hep-ph/0412079}]

\bibitem{Melosh:1974cu}
H.~J.~Melosh,
Quarks: Currents and constituents,
\href{https://journals.aps.org/prd/abstract/10.1103/PhysRevD.9.1095}
{Phys. Rev. D \textbf{9}, 1095 (1974)}.

\bibitem{Yu:2007hp}
J.~h.~Yu, B.~W.~Xiao and B.~Q.~Ma,
Space-like and time-like pion-rho transition form factors in the light-cone formalism,
\href{https://iopscience.iop.org/article/10.1088/0954-3899/34/7/021}
{J. Phys. G \textbf{34}, 1845-1860 (2007)}.
[\href{https://arxiv.org/abs/0706.2018}
{arXiv:0706.2018}]

\bibitem{Wu:2025kdc}
S.~B.~Wu, H.~J.~Tian, Y.~L.~Yang, W.~Cheng, H.~B.~Fu and T.~Zhong,
Footprint in fitting $B\to D$ vector form factor and determination for $D$-meson leading-twist LCDA,
[\href{https://arxiv.org/abs/2501.02694}
{arXiv:2501.02694}].

\bibitem{Huang:2001xb}
T.~Huang, Z.~H.~Li and X.~Y.~Wu,
Improved approach to the heavy to light form-factors in the light cone QCD sum rules,
\href{https://journals.aps.org/prd/abstract/10.1103/PhysRevD.63.094001}
{Phys. Rev. D \textbf{63}, 094001 (2001)}.

\bibitem{Ball:2007zt}
P.~Ball, V.~M.~Braun and A.~Lenz,
Twist-4 distribution amplitudes of the $K^*$ and $\phi$-mesons in QCD,
\href{https://iopscience.iop.org/article/10.1088/1126-6708/2007/08/090}
{JHEP \textbf{08}, 090 (2007)}.
[\href{https://arxiv.org/abs/0707.1201}
{arXiv:0707.1201}]

\bibitem{Fu:2018yin}
H.~B.~Fu, L.~Zeng, R.~L\"u, W.~Cheng and X.~G.~Wu,
The $D\to \rho$ semileptonic and radiative decays within the light-cone sum rules,
\href{https://link.springer.com/article/10.1140/epjc/s10052-020-7758-4}
{Eur. Phys. J. C \textbf{80}, no.3, 194 (2020)}.
[\href{https://arxiv.org/abs/1808.06412}
{arXiv:1808.06412}]

\bibitem{Hu:2024tmc}
D.~D.~Hu, X.~G.~Wu, L.~Zeng, H.~B.~Fu and T.~Zhong,
Improved light-cone harmonic oscillator model for the $\ensuremath{\phi}$-meson longitudinal leading-twist light-cone distribution amplitude and its effects to $D_s^+\to\phi\ell^+\nu_\ell,$
\href{https://journals.aps.org/prd/abstract/10.1103/PhysRevD.110.056017}
{Phys. Rev. D \textbf{110}, no.5, 056017 (2024)}.
[\href{https://arxiv.org/abs/2403.10003}
{arXiv:2403.10003}]

\bibitem{Zhong:2023cyc}
T.~Zhong, Y.~H.~Dai and H.~B.~Fu,
$\rho$-meson longitudinal leading-twist distribution amplitude revisited and the $D\to\rho$ semileptonic decay*,
\href{https://iopscience.iop.org/article/10.1088/1674-1137/ad34be}
{Chin. Phys. C \textbf{48}, no.6, 063108 (2024)}.
[\href{https://arxiv.org/abs/2308.14032}
{arXiv:2308.14032}]

\bibitem{ParticleDataGroup:2024cfk}
S.~Navas \textit{et al.} [Particle Data Group],
Review of particle physics,
\href{https://academic.oup.com/ptep/article/2022/8/083C01/6651666?login=true}
{Phys. Rev. D \textbf{110}, no.3, 030001 (2024)}.

\bibitem{Ball:2004ye}
P.~Ball and R.~Zwicky,
New results on $B \to \pi, K, \eta$ decay formfactors from light-cone sum rules,
\href{https://journals.aps.org/prd/abstract/10.1103/PhysRevD.71.014015}
{Phys. Rev. D \textbf{71}, 014015 (2005)}.
[\href{https://arxiv.org/abs/hep-ph/0406232}
{hep-ph/0406232}]

\bibitem{Huang:2013yya}
T.~Huang, T.~Zhong and X.~G.~Wu,
Determination of the pion distribution amplitude,
\href{https://journals.aps.org/prd/abstract/10.1103/PhysRevD.88.034013}
{Phys. Rev. D \textbf{88}, 034013 (2013)}.
[\href{https://arxiv.org/abs/1305.7391}
{arXiv:1305.7391}]

\bibitem{Brodsky:1981jv}
S.~J.~Brodsky, T.~Huang and G.~P.~Lepage,
Hadronic wave functions and high momentum transfer interactions in quantum chromodynamics,
Conf. Proc. C \textbf{810816}, 143-199 (1981)
SLAC-PUB-16520.

\bibitem{Wu:2010zc}
X.~G.~Wu and T.~Huang,
An Implication on the Pion Distribution Amplitude from the Pion-Photon Transition Form Factor with the New BABAR Data,
\href{https://journals.aps.org/prd/abstract/10.1103/PhysRevD.82.034024}
{Phys. Rev. D \textbf{82}, 034024 (2010)}.
[\href{https://arxiv.org/abs/1005.3359}
{arXiv:1005.3359}]

\bibitem{Wu:2011gf}
X.~G.~Wu and T.~Huang,
Constraints on the Light Pseudoscalar Meson Distribution Amplitudes from Their Meson-Photon Transition Form Factors,
\href{https://journals.aps.org/prd/abstract/10.1103/PhysRevD.84.074011}
{Phys. Rev. D \textbf{84}, 074011 (2011)}.
[\href{https://arxiv.org/abs/1106.4365}
{arXiv:1106.4365}]

\bibitem{Huang:1994dy}
T.~Huang, B.~Q.~Ma and Q.~X.~Shen,
Analysis of the pion wave function in light cone formalism,
\href{https://journals.aps.org/prd/abstract/10.1103/PhysRevD.49.1490}
{Phys. Rev. D \textbf{49}, 1490-1499 (1994)}.
[\href{https://arxiv.org/abs/hep-ph/9402285}
{hep-ph/9402285}]

\bibitem{Huang:2004su}
T.~Huang and X.~G.~Wu,
A Model for the twist-3 wave function of the Pion and its contribution to the pion form-factor,
\href{https://journals.aps.org/prd/abstract/10.1103/PhysRevD.70.093013}
{Phys. Rev. D \textbf{70}, 093013 (2004)}.
[\href{https://arxiv.org/abs/hep-ph/0408252}
{hep-ph/0408252}]

\bibitem{Wu:2005kq}
X.~G.~Wu and T.~Huang,
$\pi$ electromagnetic form-factor in the $K_T$ factorization formulae,
\href{https://www.worldscientific.com/doi/abs/10.1142/S0217751X06032277}
{Int. J. Mod. Phys. A \textbf{21}, 901-904 (2006)}.
[\href{https://arxiv.org/abs/hep-ph/0507136}
{hep-ph/0507136}]

\bibitem{Gao:2014bca}
F.~Gao, L.~Chang, Y.~X.~Liu, C.~D.~Roberts and S.~M.~Schmidt,
Parton distribution amplitudes of light vector mesons,
\href{https://journals.aps.org/prd/abstract/10.1103/PhysRevD.90.014011}
{Phys. Rev. D \textbf{90}, no.1, 014011 (2014)}.
[\href{https://arxiv.org/abs/1405.0289}
{arXiv:1405.0289}]

\bibitem{Hua:2020gnw}
J.~Hua \textit{et al.} [Lattice Parton],
Distribution Amplitudes of $K^*$ and $\phi$ at the Physical Pion Mass from Lattice QCD,
\href{https://journals.aps.org/prl/abstract/10.1103/PhysRevLett.127.062002}
{Phys. Rev. Lett. \textbf{127}, no.6, 062002 (2021)}.
[\href{https://arxiv.org/abs/2011.09788}
{arXiv:2011.09788}]

\bibitem{Tian:2024ubt}
H.~J.~Tian, H.~B.~Fu, T.~Zhong, Y.~X.~Wang and X.~G.~Wu,
Rare decay $B^+\to K^+\ell^+ \ell_-(\nu\bar{\nu})$ under the QCD sum rules approach,
\href{https://journals.aps.org/prd/abstract/10.1103/PhysRevD.111.076013}
{Phys. Rev. D \textbf{111}, no.7, 076013 (2025)}.
[\href{https://arxiv.org/abs/2411.12141}
{arXiv:2411.12141}]

\bibitem{Wang:2024oty}
Y.~X.~Wang, H.~J.~Tian, Y.~L.~Yang, T.~Zhong and H.~B.~Fu,
Prospective analysis of CKM element $|V_{cd}|$ and $D^+$-meson decay constant from leptonic decays $D^+\to \ell^+\nu$,
\href{https://www.sciencedirect.com/science/article/pii/S0370269324007986?via\%3Dihub}
{Phys. Lett. B \textbf{861}, 139240 (2025)}.
[\href{https://arxiv.org/abs/2411.10660}
{arXiv:2411.10660}]



\bibitem{Hu:2023pdl}
D.~D.~Hu, X.~G.~Wu, H.~B.~Fu, T.~Zhong, Z.~H.~Wu and L.~Zeng,
Properties of the $\eta _q$ leading-twist distribution amplitude and its effects to the $B/D^+ \rightarrow \eta ^{(\prime )}\ell ^+ \nu _\ell $ decays,
\href{https://link.springer.com/article/10.1140/epjc/s10052-023-12333-w}
{Eur. Phys. J. C \textbf{84}, no.1, 15 (2024)}.
[\href{https://arxiv.org/abs/2307.04640}
{arXiv:2307.04640}]

\bibitem{Tian:2023vbh}
H.~J.~Tian, H.~B.~Fu, T.~Zhong, X.~Luo, D.~D.~Hu and Y.~L.~Yang,
Investigating the $D_s^+ \to \pi^0 \ell^+ \nu_\ell$ decay process within the QCD sum rule approach,
\href{https://journals.aps.org/prd/abstract/10.1103/PhysRevD.108.076003}
{Phys. Rev. D \textbf{108}, no.7, 076003 (2023)}.
[\href{https://arxiv.org/abs/2306.07595}
{arXiv:2306.07595}]

\bibitem{Gill:2001jp}
J.~Gill [UKQCD],
Semileptonic decay of a heavy light pseudoscalar to a light vector meson,
\href{https://www.sciencedirect.com/science/article/abs/pii/S0920563201017248?via\%3Dihub}
{Nucl. Phys. B Proc. Suppl. \textbf{106}, 391-393 (2002)}.
[\href{https://arxiv.org/abs/hep-lat/0109035}
{arXiv:hep-lat/0109035}]

\bibitem{FOCUS:2004gfa}
J.~M.~Link \textit{et al.} [FOCUS],
New measurements of the $D_+^s \to \phi \mu^+ \nu$ form-factor ratios,
\href{https://www.sciencedirect.com/science/article/abs/pii/S0370269304002990?via\%3Dihub}
{Phys. Lett. B \textbf{586}, 183-190 (2004)}.
[\href{https://arxiv.org/abs/hep-ex/0401001}
{arXiv:hep-ex/0401001}]

\bibitem{BaBar:2006typ}
B.~Aubert \textit{et al.} [BaBar Collaboration],
Measurement of the hadronic form-factors in $D_s^+\to\phi e^{+}\nu_e$ decays,
[\href{https://arxiv.org/abs/hep-ex/0607085}
{arXiv:hep-ex/0607085}].

\bibitem{Ball:2007rt}
P.~Ball and G.~W.~Jones,
Twist-3 distribution amplitudes of $K^*$ and $\phi$ mesons,
\href{https://iopscience.iop.org/article/10.1088/1126-6708/2007/03/069}
{JHEP \textbf{03}, 069 (2007)}.
[\href{https://arxiv.org/abs/hep-ph/0702100}
{arXiv:hep-ph/0702100}]

\bibitem{BESIII:2015tql}
M.~Ablikim \textit{et al.} [BESIII Collaboration],
Study of Dynamics of $D^0 \to K^- e^+ \nu_{e}$ and $D^0\to\pi^- e^+ \nu_{e}$ Decays,
\href{https://journals.aps.org/prd/abstract/10.1103/PhysRevD.92.072012}
{Phys. Rev. D \textbf{92}, no.7, 072012 (2015)}.
[\href{https://arxiv.org/abs/1508.07560}
{arXiv:1508.07560}]

\bibitem{BESIII:2015jmz}
M.~Ablikim \textit{et al.} [BESIII Collaboration],
Study of decay dynamics and $CP$ asymmetry in $D^+ \to K^0_L e^+ \nu_e$ decay,
\href{https://journals.aps.org/prd/abstract/10.1103/PhysRevD.92.112008}
{Phys. Rev. D \textbf{92}, no.11, 112008 (2015)}.
[\href{https://arxiv.org/abs/1510.00308}
{arXiv:1510.00308}]

\bibitem{BESIII:2021anh}
M.~Ablikim \textit{et al.} [BESIII Collaboration],
Measurement of the absolute branching fractions for purely leptonic $D_s^+$ decays,
\href{https://journals.aps.org/prd/abstract/10.1103/PhysRevD.104.052009}
{Phys. Rev. D \textbf{104}, no.5, 052009 (2021)}.
[\href{https://arxiv.org/abs/2102.11734}
{arXiv:2102.11734}]


\bibitem{HeavyFlavorAveragingGroupHFLAV:2024ctg}
S.~Banerjee \textit{et al.} [Heavy Flavor Averaging Group (HFLAV)],
Averages of $b$-hadron, $c$-hadron, and $\tau$-lepton properties as of 2023,
\href{https://arxiv.org/abs/2411.18639}
{arXiv:2411.18639}.

\bibitem{Bolognani:2024cmr}
C.~Bolognani, M.~Reboud, D.~van Dyk and K.~K.~Vos,
Constraining $|V_{cs}|$ and physics beyond the Standard Model from exclusive (semi)leptonic charm decays,
\href{https://link.springer.com/article/10.1007/JHEP09(2024)099}
{JHEP \textbf{09}, 099 (2024)}.
[\href{https://arxiv.org/abs/2407.06145}
{arXiv:2407.06145}]

\bibitem{Chakraborty:2021qav}
B.~Chakraborty \textit{et al.} [(HPQCD Collaboration)\textsection{} and HPQCD],
Improved $V_{cs}$ determination using precise lattice QCD form factors for $D\to K\ell\nu$,
\href{https://journals.aps.org/prd/abstract/10.1103/PhysRevD.104.034505}
{Phys. Rev. D \textbf{104}, no.3, 034505 (2021)}.
[\href{https://arxiv.org/abs/2104.09883}
{arXiv:2104.09883}]

\bibitem{Riggio:2017zwh}
L.~Riggio, G.~Salerno and S.~Simula,
Extraction of $|V_{cd}|$ and $|V_{cs}|$ from experimental decay rates using lattice QCD $D \to \pi(K) \ell \nu$ form factors,
\href{https://link.springer.com/article/10.1140/epjc/s10052-018-5943-5}
{Eur. Phys. J. C \textbf{78}, no.6, 501 (2018)}.
[\href{https://arxiv.org/abs/1706.03657}
{arXiv:1706.03657}]

\bibitem{Belle:2013isi}
A.~Zupanc \textit{et al.} [Belle Collaboration],
Measurements of branching fractions of leptonic and hadronic $D_{s}^{+}$ meson decays and extraction of the $D_{s}^{+}$ meson decay constant,
\href{https://link.springer.com/article/10.1007/JHEP09(2013)139}
{JHEP \textbf{09}, 139 (2013)}.
[\href{https://arxiv.org/abs/1307.6240}
{arXiv:1307.6240}]

\bibitem{Charles:2004jd}
J.~Charles \textit{et al.} [CKMfitter Group],
CP violation and the CKM matrix: Assessing the impact of the asymmetric $B$ factories,
\href{https://link.springer.com/article/10.1140/epjc/s2005-02169-1}
{Eur. Phys. J. C \textbf{41}, no.1, 1-131 (2005)}.
[\href{https://arxiv.org/abs/hep-ph/0406184}
{hep-ph/0406184}]



\end{thebibliography}
\end{document}